\documentclass[fleqn,usenatbib]{mnras}

\usepackage{newtxtext,newtxmath}
\usepackage[T1]{fontenc}
\usepackage{ae,aecompl}

\usepackage{graphicx}	% Including figure files
\usepackage{amsmath}	% Advanced maths commands
\usepackage{amssymb}	% Extra maths symbols
\newcommand{\vect}[1]{\boldsymbol{#1}}
\title[Cosmic Dawn power spectrum limit with LOFAR]{The first power spectrum limit on the 21-cm signal of neutral hydrogen during the Cosmic Dawn at $z=20-25$ from LOFAR}

\author[B. K. Gehlot et al.]{B. K. Gehlot$^{1}$\thanks{E-mail: kbharatgehlot@gmail.com (BKG)}, F. G. Mertens$^{1}$, L. V. E. Koopmans$^{1}$\thanks{E-mail: koopmans@astro.rug.nl (LVEK)}, M. A. Brentjens$^{2}$, 
\newauthor S. Zaroubi$^{1,3}$, B. Ciardi$^{4}$, A. Ghosh$^{5,6}$, M. Hatef$^{1,2}$, I. T. Iliev$^{7}$, V. Jeli\'c$^{8}$,
\newauthor R. Kooistra$^{1}$, F. Krause$^{1,9}$, G. Mellema$^{10}$, M. Mevius$^{2}$, M. Mitra$^{1}$, A. R. Offringa$^{2}$,
\newauthor V. N. Pandey$^{1,2}$, A. M. Sardarabadi$^{1}$, J. Schaye$^{11}$, M. B. Silva$^{1}$, H. K. Vedantham$^{2}$,
\newauthor and S. Yatawatta$^{1,2}$
\\
$^{1}$Kapteyn Astronomical Institute, University of Groningen, PO Box 800, 9700AV Groningen, the Netherlands\\
$^{2}$ASTRON, Netherlands Institute for Radio Astronomy, Oude Hoogeveensedijk 4, 7991 PD, Dwingeloo, The Netherlands\\
$^{3}$Department of Natural Sciences, The Open University of Israel, 1 University Road, PO Box 808, Ra'anana 4353701, Israel \\
$^{4}$Max-Planck Institute for Astrophysics, Karl-Schwarzschild-Stra{\ss}e 1, 85748 Garching, Germany\\
$^{5}$Department of Physics, University of the Western Cape, Cape Town 7535, South Africa \\
$^{6}$SKA South Africa, 3rd Floor, The Park, Park Road, Pinelands, 7405 South Africa\\
$^{7}$Astronomy Centre, Department of Physics and Astronomy, Pevensey II Building, University of Sussex, Brighton BN1 9QH, U.K.\\
$^{8}$Ru{\dj}er Bo\v{s}kovi\'{c} Institute, Bijeni\v{c}ka cesta 54, 10000 Zagreb, Croatia\\
$^{9}$Department of Physics and Astronomy, University College London, Gower Place, London WC1E 6BT, U.K. \\
$^{10}$Department of Astronomy and Oskar Klein Centre for Cosmoparticle Physics, AlbaNova, Stockholm University,\\ SE-106 91 Stockholm, Sweden\\
$^{11}$Leiden Observatory, Leiden University, PO Box 9513, 2300RA Leiden, The Netherlands
}
\date{Accepted XXX. Received YYY; in original form ZZZ}

\pubyear{2019}

\begin{document}
\label{firstpage}
\pagerange{\pageref{firstpage}--\pageref{lastpage}}
\maketitle

\begin{abstract}
Observations of the redshifted 21-cm hyperfine line of neutral hydrogen from early phases of the Universe such as Cosmic Dawn and the Epoch of Reionization promise to open a new window onto the early formation of stars and galaxies. We present the first upper limits on the power spectrum of redshifted 21-cm brightness temperature fluctuations in the redshift range $z = 19.8 - 25.2$ ($54-68$ MHz frequency range) using 14 hours of data obtained with the LOFAR-Low Band Antenna (LBA) array. We also demonstrate the application of a multiple pointing calibration technique to calibrate the LOFAR-LBA dual-pointing observations centred on the North Celestial Pole and the radio galaxy 3C220.3. We observe an unexplained excess of $\sim 30-50\%$ in Stokes $I$ noise compared to Stokes $V$ for the two observed fields, which decorrelates on $\gtrsim 12$ seconds and might have a physical origin. We show that enforcing smoothness of gain errors along frequency direction during calibration reduces the additional variance in Stokes $I$ compared Stokes $V$ introduced by the calibration on sub-band level. After subtraction of smooth foregrounds, we achieve a $2\sigma$ upper limit on the 21-cm power spectrum of $\Delta_{21}^2 < (14561\,\text{mK})^2$ at $k\sim 0.038\,h\,\text{cMpc}^{-1}$ and $\Delta_{21}^2 < (14886\,\text{mK})^2$ at $k\sim 0.038 \,h\,\text{cMpc}^{-1}$ for the 3C220 and NCP fields respectively and both upper limits are consistent with each other. The upper limits for the two fields are still dominated by systematics on most $k$ modes. 
\end{abstract}

\newpage
%%%%%%%%%%%%%%%%%%%%%%%%%%%%%%%%%%%%%%%%%%%%%%%%%%

%%%%%%%%%%%%%%%%% BODY OF PAPER %%%%%%%%%%%%%%%%%%

\section{Introduction}\label{introduction}

After the Epoch of Recombination around redshift $z\sim 1100$, the Universe entered the `Dark Ages' era during which it was completely neutral and devoid of any radiation sources. During this period, small perturbations in matter density grew under gravitational instability, and matter started to accumulate in localised over-density peaks. The formation of the first luminous objects (stars and galaxies) in these overdense regions marked the beginning of the so-called Cosmic Dawn (CD) era spanning the redshift range $30 > z > 12$ \citep{pritchard2007}. X-ray and Ultraviolet radiation from the first stars and galaxies began to heat and ionize the neutral hydrogen (HI hereafter) in the surrounding Inter-Galactic Medium (IGM), starting off the Epoch of Reionization (EoR) ($12 > z > 6$) during which HI in the IGM transitioned from being fully neutral to ionized \citep{madau1997}. 

The redshifted 21-cm signal corresponding to the hyperfine transition of HI has been identified as an excellent probe of the HI distribution in the IGM during the CD and the EoR \citep{madau1997,shaver1999,furlanetto2006,pritchard2012,zaroubi2013}. A number of ongoing and upcoming experiments, such as the LOw Frequency ARray\footnote{\url{http://www.lofar.org/}}(LOFAR; \citealt{vanhaarlem2013}), the Giant Meterwave Radio Telescope\footnote{\url{http://gmrt.ncra.tifr.res.in/}}(GMRT; \citealt{paciga2011}), the Murchison Widefield Array\footnote{\url{http://www.mwatelescope.org/}}(MWA; \citealt{tingay2013,bowman2013}), the Precision Array for Probing the Epoch of Reionization\footnote{\url{http://eor.berkeley.edu/}}(PAPER; \citealt{parsons2010}), the Hydrogen Epoch of Reionization Array\footnote{\url{http://reionization.org/}}(HERA; \citealt{deboer2017}), NenuFAR\footnote{\url{https://nenufar.obs-nancay.fr/}}(New extension in Nan\c cay Upgrading loFAR; \citealt{zarka2012}), and the Square Kilometre Array\footnote{\url{http://skatelescope.org/}}(SKA; \citealt{mellema2013,koopmans2015}) are seeking to detect the brightness temperature fluctuations in the cosmological 21-cm signal using statistical methods e.g. the power spectrum. Complementary to these 21-cm power spectrum measurement experiments, several efforts such as the Experiment to Detect the Global Epoch of Reionization Signature (EDGES; \citealt{bowman2018}), the Large-aperture Experiment to Detect the Dark Ages (LEDA; \citealt{bernardi2016}), the Shaped Antenna measurement of the background RAdio Spectrum 2 (SARAS 2; \citealt{singh2017}), the Sonda Cosmol\'{o}gica de las Islas
para la Detecci\'{o}n de Hidr\'{o}geno Neutro (SCI-HI; \citealt{voytek2014}), the Probing Radio Intensity at high $z$ from Marion (PRIZM; \citealt{philip2018}), and the Netherlands-China Low frequency Explorer\footnote{\url{https://www.ru.nl/astrophysics/research/radboud-radio-lab-0/projects/netherlands-china-low-frequency-explorer-ncle/}}$^,$\footnote{\url{https://www.astron.nl/r-d-laboratory/ncle/netherlands-china-low-frequency-explorer-ncle}} (NCLE) are seeking to measure the sky-averaged spectrum of the 21-cm signal. 

At present, several instruments targeting the EoR redshifts have placed upper limits on the brightness temperature power spectrum of the redshifted 21-cm signal. \cite{paciga2013} provided the first $2\sigma$ upper limit on the brightness temperature of $\Delta_{21}^2 < (248\,\text{mK})^2$ at wavenumber $k \approx 0.5\,h\,\text{cMpc}^{-1}$ at redshift $z=8.6$ using the GMRT. \cite{beardsley2016} used MWA to set a $2\sigma$ upper limit of $\Delta_{21}^2 < (164\,\text{mK})^2$ at $k \approx 0.27\,h\,\text{cMpc}^{-1}$ at $z=7.1$. The PAPER project also provided an upper limit of $\Delta_{21}^2 < (22\,\text{mK})^2$ in the wavenumber range $0.15 \leq k \leq 0.5\,h\,\text{cMpc}^{-1}$ at $z=8.4$ \citep{ali2015}, but have recently retracted their claim due to  issues with their analysis strategy (see the erratum \citealt{ali2018}). The tightest $2\sigma$ upper limit on the 21-cm power spectrum yet is $\Delta_{21}^2 < (79.6\,\text{mK})^2$ at $k \approx 0.053\,h\,\text{cMpc}^{-1}$ in the redshift range $z= 9.6 -10.6$ and was provided by \cite{patil2017} using the LOFAR High Band Antenna (HBA) array. Instruments such as HERA, NenuFAR, and SKA-low which can potentially probe the CD redshifts are now in hardware roll-out stages (the latter is still in the development stage). \cite{ewall-wice2016} used low frequency MWA observations ($75-113$ MHz) to place an upper limit of $\Delta^2_{21} < (10^4\,\text{mK})^2$ at $k\approx 0.5$ on the power spectrum of the brightness temperature fluctuations of the 21-cm signal in the redshift range $12\lesssim z \lesssim 18$, which in most models corresponds to the epoch of X-ray heating during the CD (see e.g. \citealt{glover2003,fialkov2014,ross2017}). 

In this work, we explore, for the first time, the possibility of observing the redshifted 21-cm signal from the CD era using the LOFAR-Low Band Antenna (LBA) array which observes in the $30-90$ MHz frequency range. We use LOFAR-LBA dual pointing observations of the North Celestial Pole (NCP field hereafter) and an adjacent field centred on the 3C220.3 radio galaxy (3C220 field hereafter), which is $\sim 7^{\circ}$ away from the NCP, to study the challenges (systematic biases) in CD studies with the LOFAR-LBA and to set the first upper limits on the 21-cm brightness temperature power spectrum in the redshift range $z = 19.8 - 25.2$. We also demonstrate the application of a novel dual-pointing calibration strategy to calibrate the LOFAR-LBA data and the application of Gaussian Process Regression (GPR) as a powerful foreground removal technique in CD experiments. 

The paper is organised as follows: in Section \ref{sec:observations_preprocess}, we briefly describe the LOFAR-LBA system, the observational setup and preprocessing steps. In Section \ref{sec:calibration}, we describe the multi-beam calibration strategy to calibrate the LOFAR-LBA data. In Section \ref{sec:LBAnoise}, we assess the noise in the observed data and address the systematic biases, such as excess noise in Stokes I versus V using various statistical methods. We describe Gaussian Process Regression (GPR) in Section \ref{sec:GPR} and its application in removing residual foregrounds in LOFAR-LBA data. In Section \ref{sec:PSpec-results}, we determine the power spectra for both fields and derive upper limits on the 21-cm power spectrum in the redshift range $z = 19.8 - 25.2$. Finally, in Section \ref{sec:conclusions}, we summarise the work and discuss future prospects.

\section{Observations and preprocessing}\label{sec:observations_preprocess}

We used the LOFAR-LBA system with dual pointing setup to simultaneously observe the NCP field and the 3C220 field, which is $\sim 7^{\circ}$ away from the NCP. The NCP is the primary target field of the LOFAR-EoR KSP and has been used to set the first upper limits on the EoR power spectrum using LOFAR (see \citealt{patil2017}). The observational setup and preprocessing steps are described in the following subsections.

\subsection{LOFAR-Low Band Array}\label{subsec:lofar_lba}

The LOFAR-LBA consists of 38 stations spread across the Netherlands, providing shortest baseline lengths of $\sim 80$ m and longest baseline lengths of $\sim 100$ km. Out of these 38 stations, 24 stations (known as core stations) are spread within a core of 2 km radius, providing a densely sampled $uv$-plane. The remaining 14 stations (known as remote stations) are spread across the North-Eastern part of the Netherlands. Each LOFAR station consists of 96 low band dual-polarization dipole antennas randomly spread within an area of 81 m diameter. The voltages measured with the cross dipoles are digitised using a 200 MHz sampling clock covering the frequency range of 10-90 MHz. The digitised data is beam-formed to produce a digitally steerable station beam. At a given time, only 48 out of 96 dipoles can be combined in the beam-former. This allows a user to choose from three different station configurations in LOFAR-LBA mode viz: \texttt{LBA\_INNER} where the 48 innermost dipoles (array width $\sim 30$ m) are beam-formed, \texttt{LBA\_OUTER} where the 48 outermost dipoles (array width $\sim 81$ m) are beam-formed, and \texttt{LBA\_SPARSE} where half of the innermost 48 dipoles, plus half of the outermost 48 dipoles (array width $\sim 81$ m) are beam-formed. Each configuration results in a specific Field of View (FoV) as well as different sensitivity due to mutual coupling between the dipoles. The LOFAR-LBA system has an instantaneous bandwidth of 96 MHz. However, multiple pointings in the sky can be traded against the observable bandwidth depending on the number of pointings. In the case of two pointings, the bandwidth is reduced to 48 MHz per pointing. Readers may refer to \cite{vanhaarlem2013} for more information about the observation capabilities of LOFAR.

\subsection{Observations}\label{subsec:observations}
We use 14 hours of synthesis observation data of the NCP and the 3C220 fields, which were observed simultaneously with dual beam pointings using the \texttt{LBA\_OUTER} mode of the LOFAR-LBA system. The data were recorded during LOFAR observation Cycle 6 (ID: L557804, November 4-5, 2016). The observational details of the data are summarised in Table \ref{tab:obs_details}. The digitised data from beam-formed stations were correlated with 1 second time resolution and 3 kHz frequency resolution. The recorded data consists of 244 sub-bands for each field within the frequency range of 38-86 MHz. Each sub-band has a width of 195.3 kHz and consists of 64 channels. The recorded correlations (XX, XY, YX and YY) are stored in a Measurement Set (MS). The raw data volume for each field is $\sim18$ Terabytes and is preprocessed to reduce the data volume, as described in the next section. 

\begin{table}
	\centering
	\caption{Observational details of the data.}
	\label{tab:obs_details}
	\begin{tabular}{ll} % four columns, alignment for each		
		\hline		
		\textbf{Parameter} & \textbf{value} \\	
		\hline
		Telescope & LOFAR LBA \\
		Observation cycle and ID & Cycle 6, L557804\\ 		
		Antenna configuration & \texttt{LBA\_OUTER} \\
		Number of stations & 38 (NL stations) \\
		Observation start time (UTC) & Nov 4, 2016; 16:21:44 \\
        Number of pointings & 2 \\
		Phase centre ($\alpha,\delta$; J2000): & \\
        \quad NCP field & 00h00m00s, $+90^{\circ}00^{\prime}00^{\prime\prime}$ \\
        \quad 3C220 field & 09h39m23s, $+83^{\circ}15^{\prime}26^{\prime\prime}$\\
		Duration of observation & 14 hours   \\
        Minimum frequency & 38.08 MHz \\
        Maximum frequency & 85.54 MHz \\
		Target bandwidth & 48 MHz \\
		Primary Beam FWHM & $3.88^{\circ}$ at 60 MHz \\
		Field of View & 12 $\text{deg}^2$ at 60 MHz\\
		SEFD  & $\sim25$ kJy at 60 MHz\\		
		Polarisation & Linear X-Y   \\
		Time, frequency resolution: \\
		\quad Raw Data & 1 s, 3 kHz      \\
		\quad After flagging step 1 & 2 s, 12 kHz (archived)\\
        \quad After flagging step 2	& 2 s, 61 kHz \\
		\hline
	\end{tabular}
\end{table}

\subsection{Data selection and preprocessing}\label{subsec:preprocessing}

\begin{figure*}
\centering
\includegraphics[width=\textwidth]{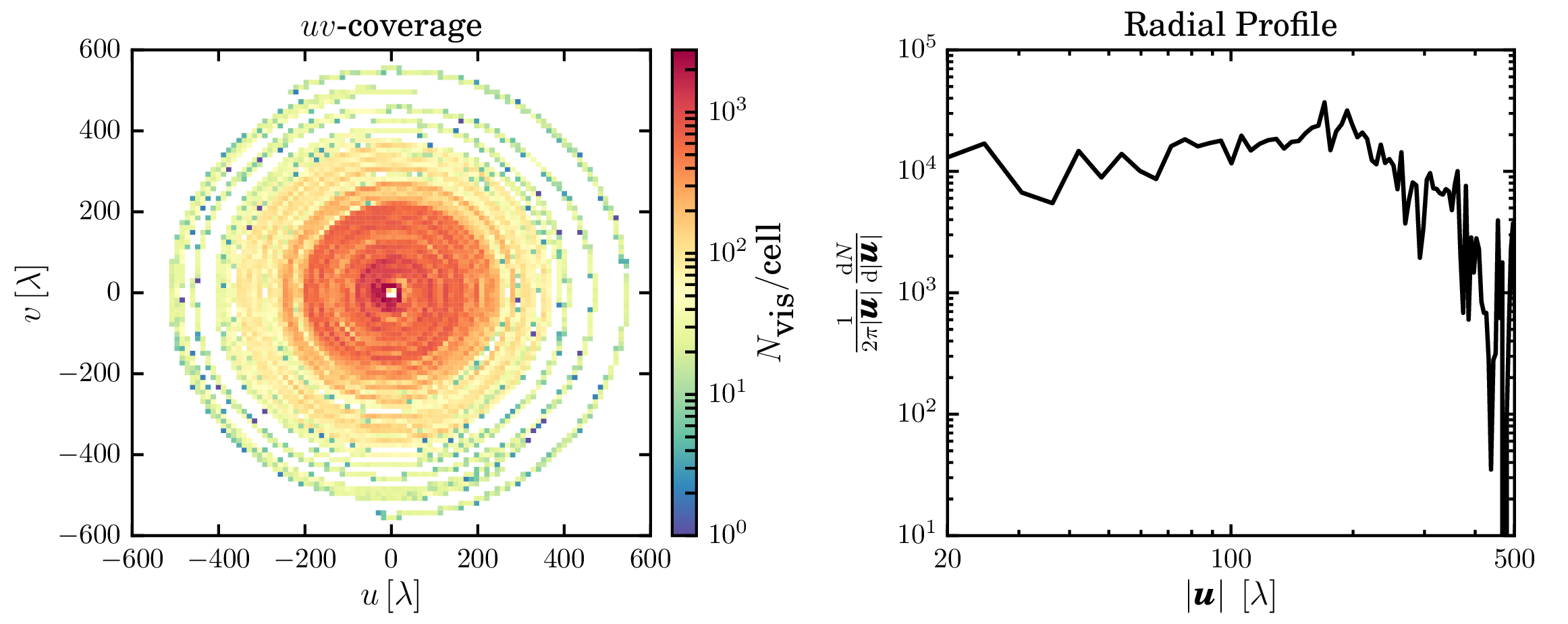}
    \caption{\textit{Left panel:} The inner region ($|\vect{u}| < 600\lambda$) of  LOFAR-LBA $uv$-coverage for the 3C220 field for 14 hour track at 60 MHz after excluding flagged visibilities. \textit{Right panel:} The radial $uv$-density $\frac{\textrm{d}N}{2\pi|\vect{u}| \textrm{d}|\vect{u}|}$ corresponding to the left panel.}
\label{fig:uvcov}
\end{figure*}

LOFAR-LBA has lower sensitivity and a relatively high RFI corruption level for frequencies above 70 MHz. Therefore, we used only 33 MHz bandwidth with the frequency range 39-72 MHz for preprocessing and further analysis. We used the standard LOFAR software pipeline (see e.g. LOFAR imaging cookbook \footnote{\url{https://www.astron.nl/radio-observatory/lofar/lofar-imaging-cookbook}}) for preprocessing the observed raw data. Processing steps include RFI-excision and averaging the data. Flagging of RFI corrupted data is performed on the highest resolution data (1 second, 3 kHz) to minimise information loss. We use the \texttt{AOFlagger} software \citep{offringa2010,offringa2012} to flag RFI corrupted data. Two channels on both edges of every sub-band were also discarded to avoid edge effects due to the polyphase filter. The remaining data were averaged in frequency and time to an intermediate-resolution of 12 kHz and 2 seconds, resulting in 15 channels per sub-band. This intermediate resolution data is archived for future use. To reduce the data volume further, it was averaged in frequency to 61 kHz and the auto-correlations were also flagged. The resulting data consists of 3 channels of 61 kHz each per sub-band and has a time resolution of 2 seconds. We flagged the remote station RS503LBA in all sub-bands for both fields because of its proximity to a windmill, which causes strong RFI in the visibilities of the station. We also observed that CS302LBA had poor gain upon inspecting the visibilities and flagged it for both fields. The flagging and averaging were performed separately on both 3C220 and NCP field datasets, although some correlation is obviously expected. Figure \ref{fig:uvcov} shows LOFAR-LBA  $uv$-coverage (the inner region with $|\vect{u}|<600\lambda$) for the 3C220 field pointing for 14 hours track at 60 MHz after exclusion of flagged visibilities and the radial profile of the $uv$-coverage. 

\section{Calibration Scheme}\label{sec:calibration}

The visibilities recorded by LOFAR are corrupted by the instrumental (complex station gains, primary beam, instrumental bandpass, clock-drift etc.) and environmental (ionosphere) factors. Calibration of the LOFAR-LBA system involves estimating the errors that corrupt the measured visibilities, and to obtain an accurate estimate of the true visibilities from observed data. Calibration of LOFAR-LBA data involves two major steps: (a) Direction Independent (DI) calibration and, (b) Direction Dependent (DD) calibration. DI calibration involves estimation of a single instrumental gain (represented by a complex $2\times 2$ Jones matrix) for each beam-formed station, and DD calibration accounts for the direction dependent errors arising from wave propagation effects through the ionosphere and the primary beam. We use \texttt{SAGECal-CO}\footnote{\url{http://sagecal.sourceforge.net/}} to perform the major calibration steps. \texttt{SAGECal-CO} performs calibration in a distributed way using consensus optimisation \citep{boyd2011}, which is an effective way to improve the quality of the calibration of radio interferometric data. In \texttt{SAGECal-CO}, the calibration problem is transformed into consensus optimisation by adding frequency smoothness of systematic errors as a constraint. It uses an Alternating Direction Method of Multipliers (ADMM) algorithm to reach convergence. Readers may refer to \cite{yatawatta2015,yatawatta2016,yatawatta2017,yatawatta2018} for a detailed description of the \texttt{SAGECal-CO} algorithm and its capabilities.
\\
Upon inspection of the raw visibilities, we observed that Cas\,A ($\sim 30^{\circ}$ away from NCP) and Cyg\,A ($\sim 50^{\circ}$ away from the NCP) superpose significant side-lobes onto both fields. It is crucial to subtract these sources before performing DI calibration to avoid errors due to these side-lobes. We use DD-calibration in \texttt{SAGECal-CO} to subtract Cas\,A and Cyg\,A. \cite{gehlot2018} (G18 hereafter) showed that the residuals after subtraction of bright sources such Cas\,A and Cyg\,A are significant as well as incoherent over timescales of a few minutes, depending on the strength of ionospheric scintillations. Therefore, we use a solution time and frequency interval of 30 seconds and 61 kHz to subtract Cas\,A and Cyg\,A, which is optimised to incorporate ionospheric effects while maintaining a decent signal-to-noise ratio ($\gtrsim 10$) for the given solution interval. We use the Cas\,A and Cyg\,A shapelet models \footnote{Cas\,A and Cyg\,A models were derived from wide-band LOFAR-LBA and HBA observations of Cas\,A and Cyg\,A. Each source has about 200 components (shapelets and point). See \cite{yatawatta2011} for more details.} as an input model for calibration and subtraction. The subtraction was performed individually on both fields. 

The two fields, 3C220 and NCP, given their different pointings and gain solutions, have slightly different morphologies. The 3C220 field consists of a reasonably bright source located at the phase centre (the 3C220.3 radio galaxy with a flux of $\sim 38$ Jy at 74 MHz \citep{cohen2007}) which can be utilised as a bandpass calibrator, making calibration of the 3C220 field fairly straightforward. However, the NCP field does not have such relatively bright sources near the phase centre, which makes it more difficult to calibrate the field. Therefore, we adopt a calibration strategy where we calibrate the 3C220 field first and then use the output calibration products to calibrate the NCP field, given that the bandpass calibration solutions should be similar between the fields because of the same electronics, and that any effect of the beam should be spectrally smooth near the phase centre. A similar technique to calibrate the LOFAR-LBA data to study the ionospheric effects is shown in de Gasperin et al. (in preparation) and \cite{degasperin2018}. Similar types of calibration strategies are more common in radio survey experiments, although in those cases it is often required to switch between sources in time. The calibration settings (e.g. solution interval, frequency resolution, ADMM iterations, regularization factor) for the two fields are chosen to account for any rapidly varying effects in time and frequency such as the ionosphere while maintaining a reasonable signal to noise ratio per solution interval. Most of these settings are decided on the basis of the analysis and lessons learned in G18 as well as the analysis of the LOFAR-EoR data (see e.g. \citealt{patil2017}).    

\begin{figure*}
\centering
\includegraphics[width=\textwidth]{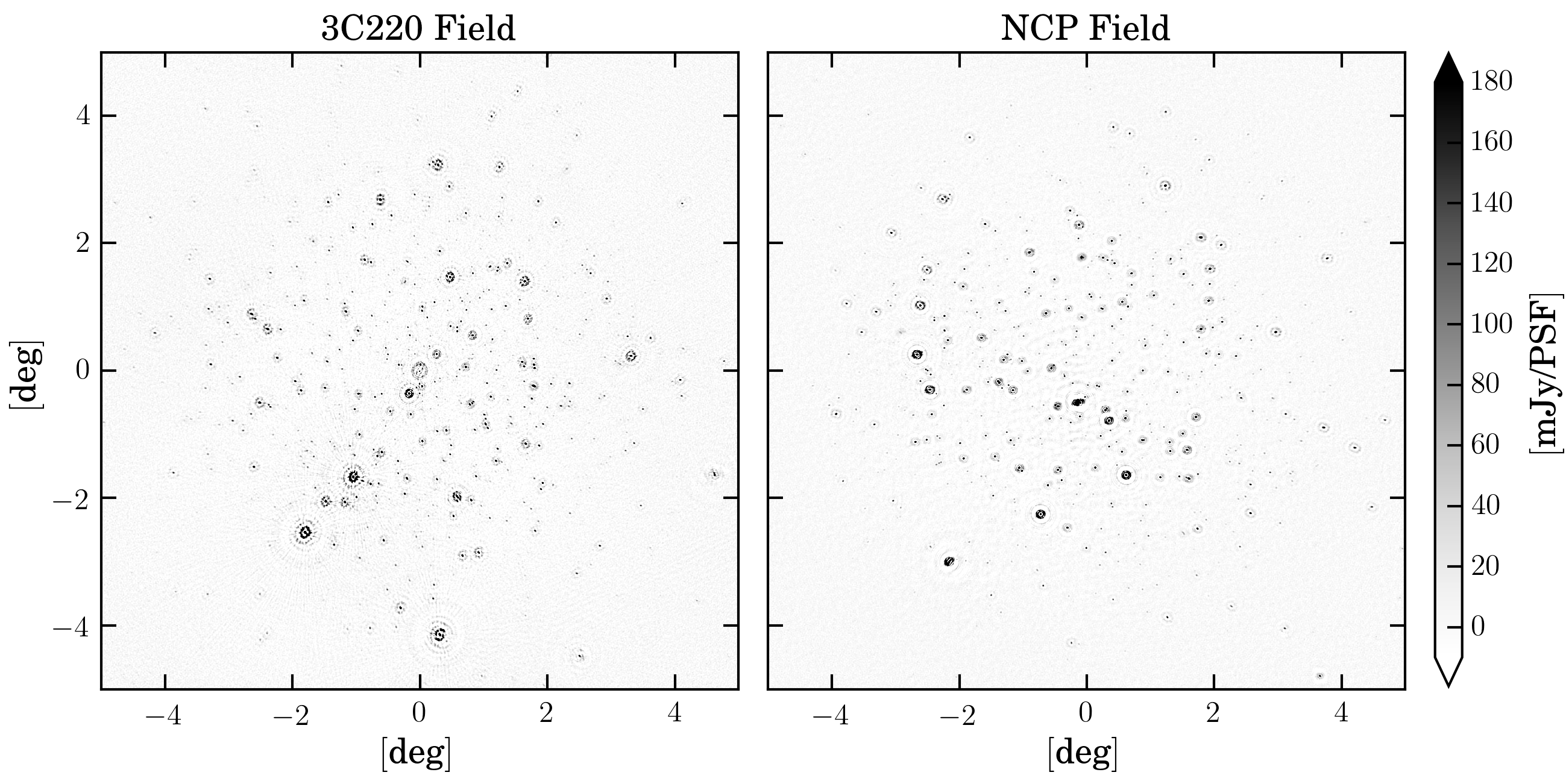}
    \caption{Left and right panels show Stokes $I$ continuum `dirty' images ($39-72$ MHz) of the 3C220 and NCP field respectively, after DI calibration. The images are not cleaned, and were produced using $\leq2000\lambda$ baselines with the Briggs -0.1 weighting scheme. The observed image rms is $\sigma_{\text{rms}}\sim 7$ mJy for the 3C220 field and $\sim 5.5$ mJy for the NCP field respectively. These values are still $\sim 10$ times higher than the expected rms ($\sim 0.7$ mJy) calculated using SEFD (System Equivalent Flux Density) estimates for LOFAR-LBA. The values of $\sigma_{\text{rms}}$ can be calculated from SEFD using the relation: $\sigma_{\text{rms}} =  \text{SEFD}/\sqrt{2N(N-1)\Delta \nu \Delta t}$, where SEFD  $\sim 30$ kJy at 55 MHz, $N = 29$ (corresponding to $2000\lambda$ baseline range), $\Delta\nu = 33$ MHz, $\Delta t = 0.9\times 14$ hours (assuming flagged data at $10\%$ level).} 
\label{fig:continuum}
\end{figure*} 

\subsection{Calibrating the 3C220 field}\label{subsec:3C220cal}
To calibrate the 3C220 field, we use a calibration strategy similar to that discussed in G18. The 3C220.3 radio galaxy is a double-lobed source of $\sim 8$ arcsec extent, but it is unresolved with the LOFAR-LBA array which has a maximum resolution of $\sim 15$ arcsec. Therefore, we use a single point source  representing 3C220.3 with $38$ Jy flux at 74 MHz \citep{cohen2007} and a spectral index of -0.8 as a starting model for DI calibration. The major steps involved in the calibration of the 3C220 field are as follows:

\begin{enumerate}

\item Calibrate the raw visibilities using the 3C220.3 point source model in \texttt{NDPPP} \footnote{\url{http://www.lofar.org/operations/doku.php?id=public:user_software:ndppp}} to obtain the station gain solutions with 30 seconds and 61 kHz calibration solution intervals and subsequently apply them to the data. This step is performed separately for each sub-band (without consensus optimisation). We include the primary beam\footnote{Current LBA primary beam models are based on Electro-Magnetic (EM) simulations of the LOFAR-LBA dipoles (private communication with LOFAR Radio Observatory).} in the calibration step in \texttt{NDPPP}. Note that the LOFAR-LBA beam model has only been implemented in \texttt{NDPPP} at present. Hence, it is utilised for primary DI calibration for both fields. Note that we do not exclude any baselines during DI calibration steps for both fields.

\item Deconvolve (clean) and image the calibrated visibilities using the \texttt{WSClean} package \citep{offringa2014} with the following settings: cleaning threshold = $0.5\sigma$, weighting scheme = uniform, imaging baseline range = $0-5000\lambda$, 4th order linear polynomial\footnote{Using log polynomials to fit source spectra is unstable in \texttt{WSClean}. Therefore, we use an ordinary 4th order linear polynomial to fit source spectra. However, \texttt{SAGECal-CO} is only compatible with log polynomials. Therefore, we separately fit the source spectra with a 3rd order log-polynomial to make it compatible with \texttt{SAGECal-CO}.} for fitting the source spectrum over 15 points which correspond to averaged flux over 2.2 MHz bands spread within 33 MHz bandwidth. The cleaning parameters are chosen such that the modelled sources with lowest flux values are still a factor of few above the confusion limit at 60\,MHz ($\sigma_c \sim 10 \text{mJy/beam}^{-1}$, see \citealt{vanhaarlem2013} for calculation of $\sigma_c$). Since we do not apply the primary beam correction during imaging, the source fluxes are apparent and their spectra also possess the primary beam variations which are less smooth compared to the intrinsic source spectra. Using a 4th order polynomial for spectral fitting easily captures these beam variations compared to a lower order polynomial and facilitates better source subtraction. Step (i) is repeated once more using the clean model of 3C220.3 obtained in step (ii) and deconvolution is performed to obtain a more accurate 3C220.3 clean model. Further iterations were not required as the model converged.

\item Use \texttt{SAGECal-CO} to perform DI calibration of raw visibilities and subtract 3C220.3 using consensus optimisation (7 iterations and regularization factor of 5) over a 33 MHz frequency range. We provide the final clean model of 3C220.3 obtained after step (ii) as input to \texttt{SAGECal-CO} and use a calibration solution interval of 30 seconds and 183.1 kHz. The obtained gain solutions are subsequently applied to the residual visibilities.

\item Repeat the deconvolution with the same settings (but with lower clean-mask = $4\sigma$) in step (ii) to clean and image the residual visibilities after step (iii). The output clean model of the radio sources in the field contains 1270 components (points plus Gaussians) with flux $> 40$ mJy at 55 MHz. We repeated Step (iii) with this updated sky-model to perform DI-calibration and subtraction of 3C220.3 from the visibilities. Using a more complete sky-model in DI calibration allows for the mitigation of calibration errors due to unmodeled sources and produces accurately calibrated visibilities. The gain solutions obtained after this step are later utilised in the calibration of the NCP field.

\item Use DD-calibration with \texttt{SAGECal-CO} to subtract the clean-model obtained in step (iv). \texttt{SAGECal-CO} accounts for DD errors by obtaining the gain solutions in multiple directions. It subtracts the sources in each direction by multiplying the obtained gain solutions with the predicted visibilities and subtracting the product from the observed visibilities. We divide the 1270 components into 4 clusters using the weighted K-means clustering algorithm \citep{kazemi2013a} and use the centres of these clusters as four different directions. This algorithm operates on a unit sphere and the corresponding weights are determined by the source flux. The algorithm creates clusters such that the cluster size is minimized while maintaining similar net flux in each cluster to ensure sufficient signal to noise ratio for each cluster. We use a gain solution interval of 20 minutes and 183.1 kHz and 20 ADMM iterations for each gain solution while keeping the same regularization factor of $\rho=5$ \citep{yatawatta2016} as in DI calibration. We discard the baselines $\leq 200\lambda$ in the DD-calibration. These excluded baselines ($<200\lambda$) are used for further analyses and power spectrum estimation. Using a baseline cut avoids any bias due to unmodeled diffuse emission on shorter baselines excluded from calibration (see e.g. \citealt{patil2016,barry2016,ewall-wice2017,gehlot2018} for more details). It also mitigates the suppression of the 21-cm signal caused by the use of an incomplete sky-model in the calibration, as shown in \cite{patil2017} and \cite{sardarabadi2018}, and we will test this further in future. However, the exclusion of short baselines from the calibration also has a drawback that longer baselines are prone to calibration errors. These errors, when applied to excluded baselines, cause the foregrounds to leak into the ``EoR-window" on corresponding baselines \citep{barry2016,patil2016}. Using the smoothness of gain solutions as a constraint in the calibration largely mitigates this effect \citep{sardarabadi2018}. An optimal baseline selection criteria for calibration which accounts for these effects itself requires fairly rigorous analysis and is beyond the scope of this paper. The choice of $200\lambda$ baseline cut comes from the reason that the radial profile of the $uv$-coverage (see right panel of figure \ref{fig:uvcov}) is relatively flat within $20\lambda \leq |\vect{u}| \leq 200\lambda$ range and drops for longer baselines. This is an optimal choice for power spectrum estimation because of relatively uniform $uv$-coverage.

\item Image the residual visibilities in step (v) with \texttt{WSClean}. We used the following settings: weighting scheme = natural, pixel size = 3 arcmin, Image dimensions = $300\times 300$ pixels, imaging baselines = $15 - 200 \lambda$. Note that we do not deconvolve the final residual images. The output Stokes $I$, $V$ and Point Spread Function (PSF) images were stored for further analysis. The left panel of figure \ref{fig:continuum} shows the dirty continuum image of the 3C220 field after DI calibration where the 3C220.3 has been subtracted.

\end{enumerate}
 
 \begin{figure*}
\centering
\includegraphics[width=\textwidth]{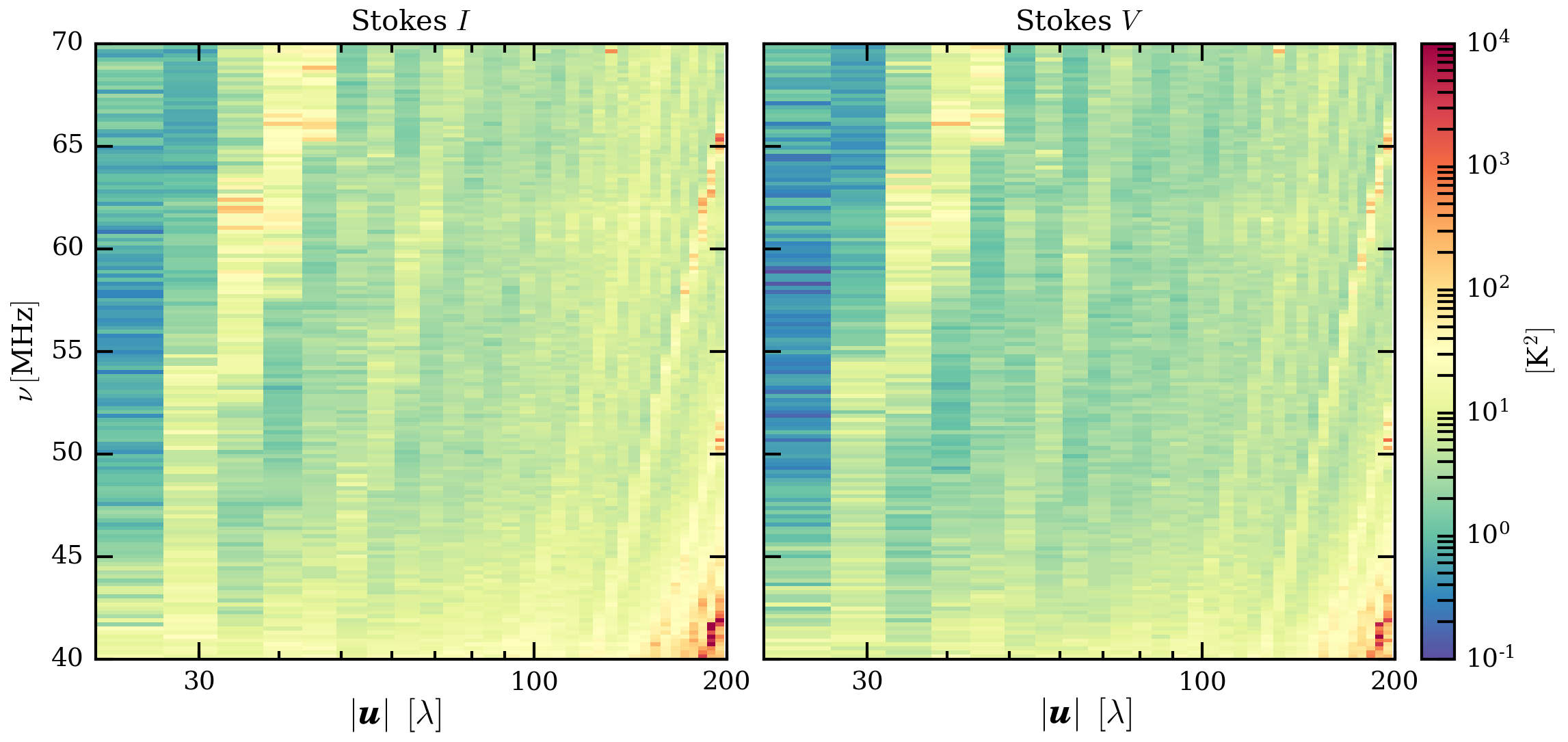}
    \caption{Stokes $I$ ($P_{\Delta_t I}(|\vect{u}|,\nu)$) and $V$ ($P_{\Delta_t V}(|\vect{u}|,\nu)$) noise spectra for the 3C220 field. Left and right panels correspond to Stokes $I$ and $V$ respectively.} 
\label{fig:noise_PS_IV}
\end{figure*} 
 
\subsection{Calibrating the NCP field}\label{subsec:NCPcal}

The absence of very bright sources makes the NCP field more difficult to calibrate using the strategy we employed for the 3C220 field. Therefore, we utilise a different approach. The NCP field consists of a moderately bright source (3C061.1) which lies at the edge of the primary beam causing the source to exhibit peculiar behaviour in its gain solutions. We, therefore, subtract 3C061.1 from the raw visibilities using DD-calibration with \texttt{SAGECal-CO} with the same settings as we employed for the Cas\,A and Cyg\,A subtractions. The 3C061.1 input model is adapted from the intrinsic model of 3C061.1 (points + shapelets, at 150 MHz) used in the LOFAR-EoR data processing pipeline (see e.g. \citealt{patil2017}). The fluxes in the model were scaled properly to match the flux values quoted in \cite{laing1980} and \cite{hales1995}. After subtraction of 3C061.1, visibilities were calibrated using the following steps:

\begin{enumerate}
\item Apply the DI gain solution amplitudes of the 3C220 field obtained in step (iv) in section \ref{subsec:3C220cal} to the NCP field visibilities to set the amplitude scale.

\item Deconvolve (clean) and image the resulting visibilities using \texttt{WSClean} with the following settings: cleaning threshold = $0.5\sigma$, weighting scheme = uniform, imaging baseline range = $0-2000\lambda$, 2nd order polynomial for fitting the source spectrum over 5 points which correspond to an averaged flux over 6.6 MHz bands spread over 33 MHz. 

\item Perform DI calibration of the visibilities with \texttt{SAGECal-CO} using consensus optimisation (with same settings as in DI calibration of the 3C220 field) over the 33 MHz frequency range. The clean model obtained in step (ii) is provided as input. We use a calibration solution interval of 10 minutes and 183.1 kHz. The obtained gain solutions are subsequently applied to the visibilities. We repeat steps (ii) and (iii) in a self-cal loop with 3 iterations. The final clean model after 3 self-cal iterations contains 1470 components (points plus Gaussians) with flux $> 40$ mJy at 55 MHz. 

\item Perform phase calibration using \texttt{NDPPP} on the visibilities obtained after step (i). We use the final clean model obtained after step (iii) as input and choose 30 seconds, 183.1 kHz as the calibration solution interval.

\item Use DD-calibration with \texttt{SAGECal-CO} to subtract the clean-model obtained in step (iii). We divide 1470 components into three clusters representing three directions (which represent three non-overlapping regions within the primary beam) using the weighted K-means clustering algorithm (same as in step (v) of section \ref{subsec:3C220cal}). We use a gain solution interval of 20 minutes and 183.1 kHz and 20 ADMM iterations for each gain solution. We discard the baselines $\leq 200\lambda$ to avoid errors due to unmodeled diffuse emission on shorter baselines and to avoid signal suppression. 

\item Image the residual visibilities in step (v) with \texttt{WSClean} using the following settings: weighting scheme = natural, pixel-size = 3 arcmin, Image dimensions = $300\times 300$ pixels, imaging baselines = $15 - 200 \lambda$. The output Stokes $I$, $V$ and PSF images were stored for further analysis. The right panel of figure \ref{fig:continuum} shows the dirty continuum image of the NCP field after DI calibration.

\end{enumerate}

We only use the beam model during the DI-calibration and image deconvolution steps, and we do not correct the residual images for the primary beam. Also, we do not  analyse the Stokes $Q$ and $U$ data. The latter is mainly because we do not include any polarised (compact or diffuse) emission in sky-model used for the calibration. Any unmodeled emission in Stokes $Q$ and $U$ can essentially bias the calibration. The currently utilised calibration scheme is defined such that only the Stokes $I$ and $V$ are constrained by the sky-model, whereas, the Stokes $Q$, $U$ have the freedom to rotate. Moreover, in the RM-synthesis analysis in G18, we did not observe any signature of the polarised emission in RM-space, suggesting the absence of significant polarised emission at these low frequencies. Because the Rotation Measure scales as $\lambda^2$, any polarised emission at low frequencies (40-70\,MHz) is essentially depolarised by the intervening magneto-ionic medium and the rapidly varying ionosphere. However, in G18, we observed strong polarization leakage of the bright off-axis source Cas\,A from Stokes $I$ to $Q$, $U$, and also in Stokes $V$ but at a much weaker level compared to Stokes $Q$ and $U$. This effect is mitigated by subtraction of Cas\,A and Cyg\,A during using DD-calibration at higher time resolution (30 seconds) and is already accounted for in the current analysis. The leakage of (partly) polarised foregrounds to Stokes $I$, if any, is expected to be significantly lower than the current noise level and currently does not affect our analysis. 

At this point, we have residual data cubes that are DI calibrated and where the sky model has been subtracted using their DD gain solutions. These residual cubes form the input for subsequent analyses. In the following sections, we will discuss these analyses.

\begin{figure}
\centering
\includegraphics[width=\columnwidth]{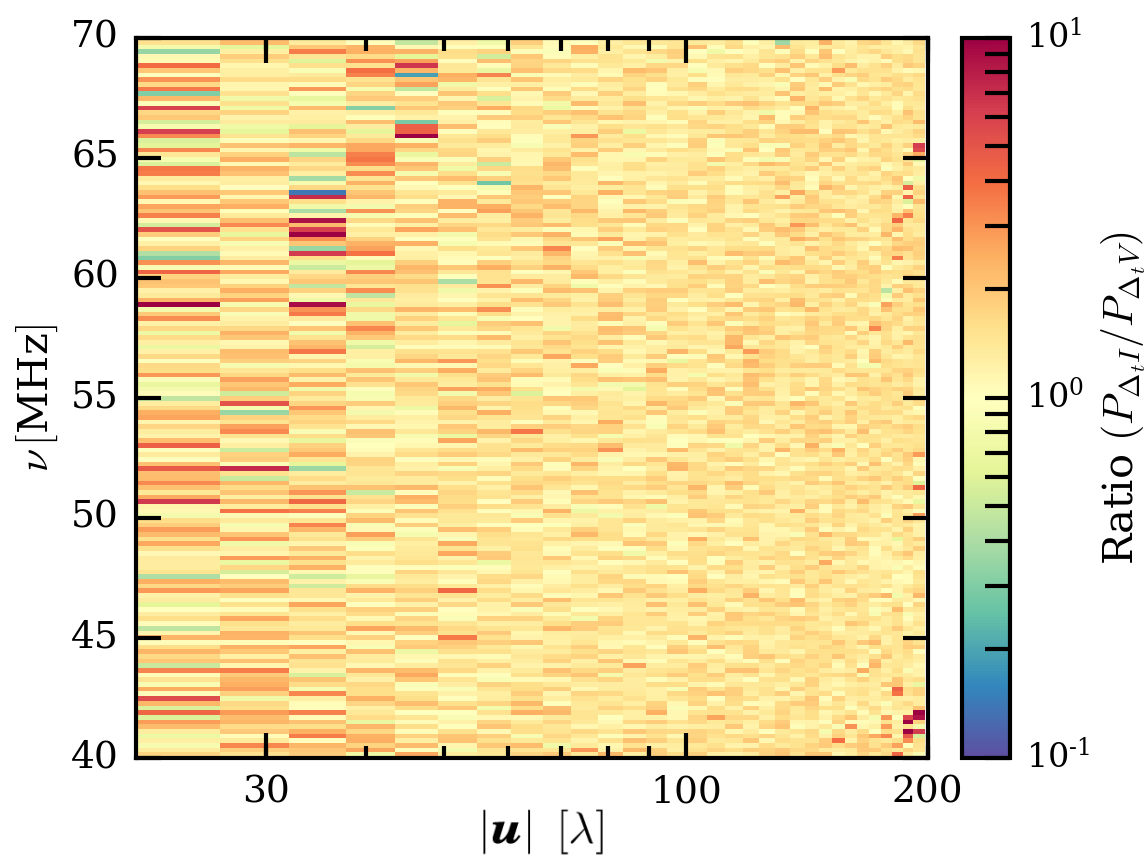}
    \caption{The ratio $P_{\Delta_t I} / P_{\Delta_t V}$ of the noise spectra shown in figure \ref{fig:noise_PS_IV}. The ratio is flat except for a few outliers at shorter baselines.} 
\label{fig:Ratio_noise_PS_IV}
\end{figure} 

\begin{figure*}
\centering
\includegraphics[width=\textwidth]{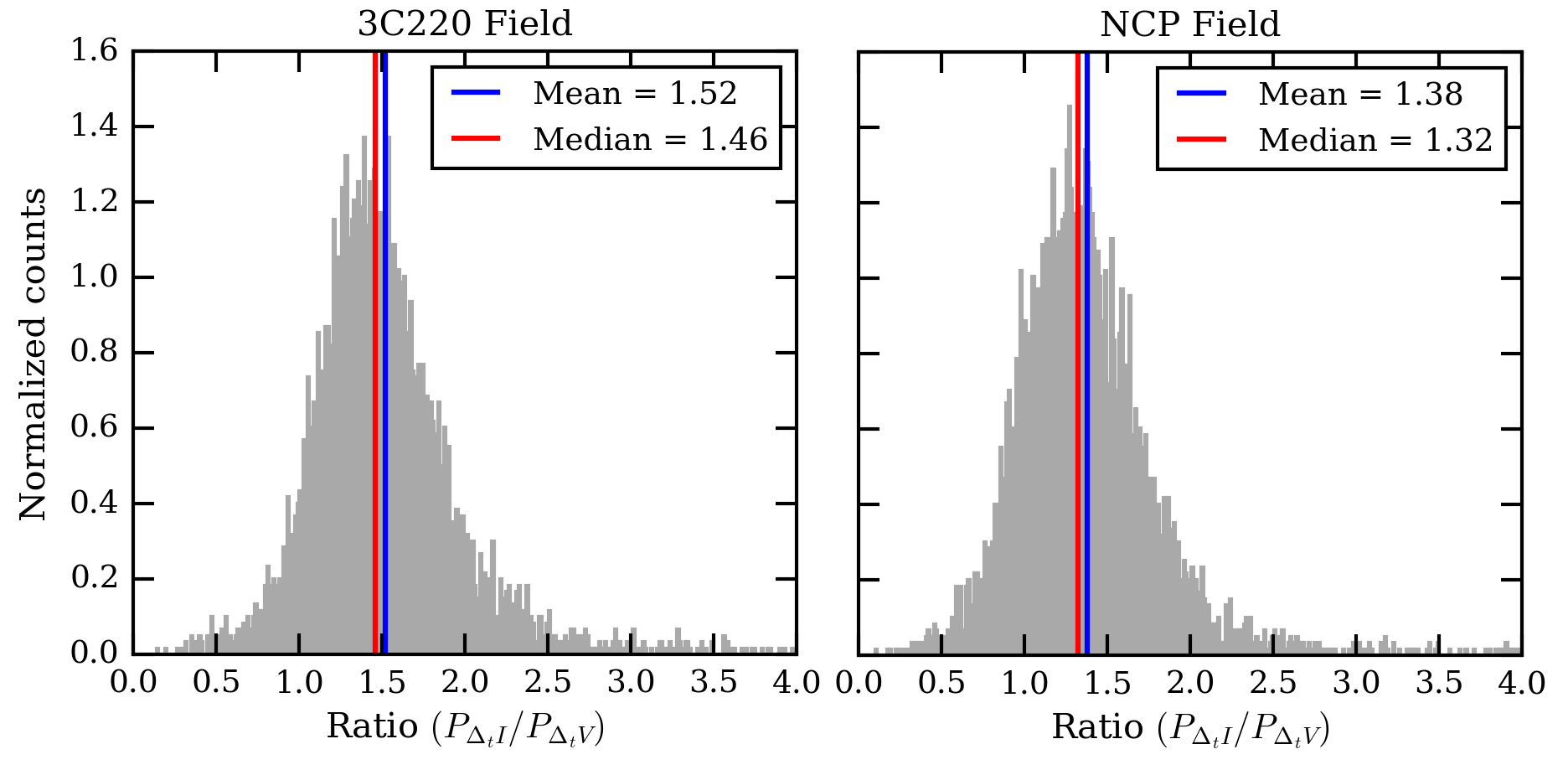}
    \caption{The left and the right panels show normalised histograms of the distribution of the ratio values $P_{\Delta_t I} / P_{\Delta_t V}$ for the 3C220 and the NCP fields, respectively. The red and blue vertical lines represent the median and the mean of the distribution respectively. For the 3C220 field, the distribution has a median value of $1.46$ and a mean value of $1.52$. Similarly, the median and the mean values for the NCP field are $1.32$ and $1.38$ respectively.} 
\label{fig:Hist_noise-ratio}
\end{figure*} 

\section{Noise statistics in LOFAR-LBA}\label{sec:LBAnoise}

Current estimates of the average Signal Equivalent Flux Density (SEFD) per station of the LOFAR-LBA array are derived from the observations of bright sources at zenith. However, the SEFD of LOFAR varies as a function of angle from the zenith. Therefore, using zenith SEFD estimates to derive the noise on the visibilities and rms in the images (also noise power spectra) typically underestimates the SEFD for the fields away from the zenith. To avoid this bias, we estimate the noise and hence the noise spectrum (in baseline-frequency space) for the 3C220 field from the visibilities. A standard method to estimate the noise on visibilities is to subtract the un-gridded visibilities corresponding for two contiguous time-steps at the highest time resolution. However, this method is not feasible for large LOFAR-LBA datasets ($\sim18$ TB per dataset) because of a large number of baselines and time-steps. Therefore, we use an alternative approach where we estimate the noise spectrum from the gridded visibilities (see e.g \citealt{jacobs2016,beardsley2016,ewall-wice2016}). We expect that these two approaches become equivalent to each other for datasets with a large number of time-samples and leave additional comparison tests between the two approaches for future analyses. 

We split the DI-calibrated visibilities of the 3C220 field into even and odd samplings with 12 seconds cadence such that these samplings are interleaved in time. Note that for the baseline range $20\lambda \leq |\vect{u}|\leq 200\lambda$ which we probe in our analysis, the sky and the PSF do not vary over a 12 second interval. Also, any sky leakage over 12 seconds will appear as a wedge in the cylindrically averaged power spectrum, which we do not observe in the analyses (shown in later sections). Moreover, we expect the system to be coherent over 12 seconds and only ionospheric effects are expected to change. We image these even and odd samplings using \texttt{WSClean} with the `natural' weighting scheme. We Fourier Transform (FT) the even and odd image cubes and properly scale visibilities in each $uv$-cell with corresponding sampling density to remove the effect of gridding weights during imaging. We calculate the azimuthally averaged (spatial) power spectrum of the difference as $P_{\Delta_t I}(|\vect{u}|,\nu) \equiv \langle \Delta_{t}\tilde{I} \rangle^2 = \langle \tilde{I}_{\text{even}} - \tilde{I}_{\text{odd}} \rangle^2 / 2$, where $\tilde{I}_{\text{even}}$ and $\tilde{I}_{\text{odd}}$ are the Fourier transforms of the  even and odd Stokes $I$ image cubes respectively, $\vect{u} = (u,v)$ is the baseline vector (in units of wavelength) in the $uv$-plane and $|\vect{u}| = \sqrt{u^2 + v^2}$ and $\nu$ is the frequency. Similarly, $P_{\Delta_t V}(|\vect{u}|,\nu)\equiv \langle \Delta_{t}\tilde{V} \rangle^2$ is determined using the even and odd Stokes $V$ image cubes.

\subsection{Physical Excess Noise}\label{subsec:ExcessNoise}
Figure \ref{fig:noise_PS_IV} shows $P_{\Delta_t I}$ and $P_{\Delta_t V}$ for the $20-200 \lambda$ baseline range for the 3C220 field. We observe that both $P_{\Delta_t I}$ and $P_{\Delta_t V}$ spectra are relatively flat. The bright tilted streaks are a consequence of varying $uv$-density as a function of baseline length in LOFAR-LBA. We compare $P_{\Delta_t I}$ and $P_{\Delta_t V}$ by calculating their ratio. Figure \ref{fig:Ratio_noise_PS_IV} shows the ratio $P_{\Delta_t I} / P_{\Delta_t V}$ of the spectra shown in figure \ref{fig:noise_PS_IV}. We observe that the ratio is remarkably flat, except for a few outliers at shorter baselines ($\leq 40 \lambda$). Most of these outliers are also coincident with baselines where $uv$-density is relatively low. These outliers might arise due to imperfect calibration and slight differences in flagging of RFI affected baselines post calibration along with $uv$-density variations. Ideally, if the noise properties of Stokes $I$ and $V$ are statistically identical and if the sky and the PSF do not change over a 12 seconds interval, $P_{\Delta_{t} I}$ and $P_{\Delta_{t} V}$ are expected to be identical assuming that the sky has a negligible circular polarised emission component and Stokes $V$ is virtually empty. However, we observe excess power in Stokes $I$ compared to Stokes $V$, which is largely constant over the $20 - 200\lambda$ baseline range and over the 30 MHz bandwidth. Although the power in both Stokes $I$ and $V$ varies slightly with increasing baseline length, the ratio remains constant, suggesting that this slight variation is a result of varying $uv$-density. Most correlations that are spectrally smooth, e.g. due to intrinsic foregrounds, instrumental mode-mixing and ionosphere, appear as a ``wedge" in the two-dimensional power spectrum. Whereas systematic effects with specific spectral structure e.g. cable reflections may appear as a distinct feature above the ``wedge". However, only those effects that are non-smooth in frequency or possess noise-like behaviour affect most scales in the two-dimensional power spectrum. In later sections (see section \ref{subsec:2Dcompare_with_noise}) we show that the corresponding 2D power spectra for Stokes $I$ and $V$ noise estimates are featureless and devoid of any ``wedge" like structure or other distinct features corresponding to systematic effects. Hence this physical excess noise, for all practical purposes, is treated as additional white noise in Stokes $I$ that is seemingly uncorrelated in frequency and remains more or less the same for different baseline lengths. 

The left panel of figure \ref{fig:Hist_noise-ratio} shows the normalised histogram of the distribution of $P_{\Delta_t I} / P_{\Delta_t V}$ values for the 3C220 field. The distribution has a median value of $1.46$ and a mean value of $1.54$, with most values lying within the range $1-2$. The noise spectra and their ratio for the NCP field also exhibit similar behaviour as the 3C220 field that the ratio $P_{\Delta_t I} / P_{\Delta_t V}$ is flat in frequency-baseline space. However, the distribution of the ratio values (see right panel of figure \ref{fig:Hist_noise-ratio}) has a slightly lower median and mean values of $1.32$ and $1.38$ respectively. The cause of this excess power in $P_{\Delta_t I}$ is still unknown, but it is higher for the 3C220 field which has a bright source at the centre, compared to the NCP field which is devoid of relatively bright sources. We are currently investigating the cause of this excess, but given that the excess is different for the two fields, ionospheric or interplanetary scintillation noise might be a probable reason for this excess, although the rapid decorrelation with frequency remains unexplained.

\begin{figure}
\centering
\includegraphics[width=\columnwidth]{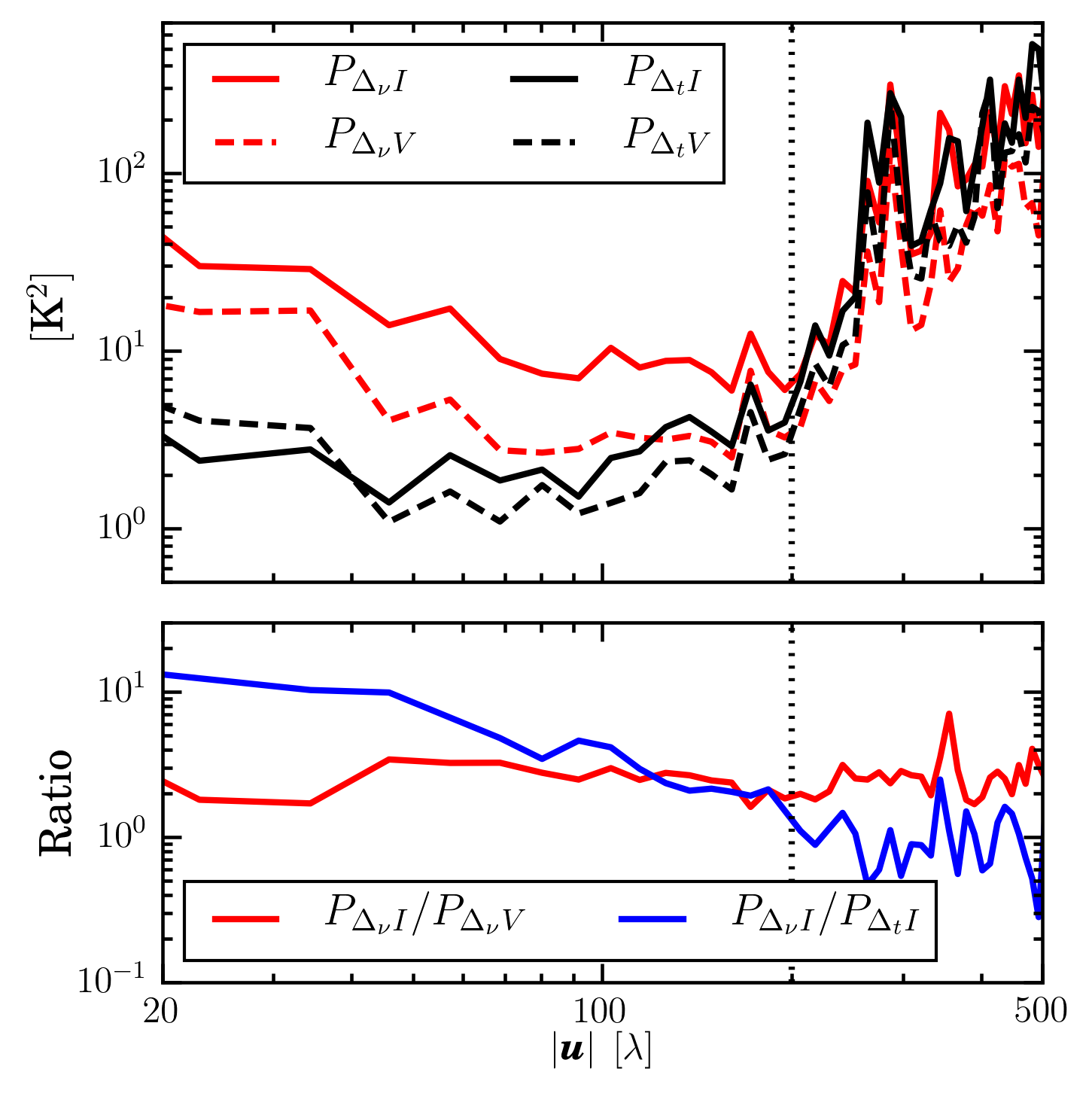}
    \caption{\textit{Top panel:} The differential Stokes $I$ and $V$ power spectra calculated using residual images of the 3C220 field. The solid red curve corresponds to $P_{\Delta_{\nu} I}$ and the dashed red curve corresponds to $P_{\Delta_{\nu} V}$. The solid and dashed black curves correspond to $P_{\Delta_{t} I}$ and $P_{\Delta_{t} V}$ respectively, at $\nu = 59.95$ MHz. \textit{Bottom panel:} The ratio $P_{\Delta_{\nu} I}/P_{\Delta_{\nu} V}$ (red curve) and the ratio $P_{\Delta_{\nu} I}/P_{\Delta_{t} I}$ (blue curve). The dotted vertical line shows the location of the $200\lambda$ baseline cut.} 
\label{fig:dPS_IV}
\end{figure}

\subsection{Variance at the inter-sub-band level}\label{subsec:excess_compare}

We use the azimuthally averaged power spectrum of the difference of Stokes $I$ and $V$ images between two contiguous sub-bands (differential power spectrum) to study the behaviour of variance at the inter-sub-band level (see e.g. \citealt{patil2016,gehlot2018}). This method is based on the assumption that Stokes $I$ images are composed of total sky signal convolved with the PSF plus additive noise. Assuming that the sky signal, which is smooth in frequency does not change\footnote{ For a spectral index of $-2.55$, sky brightness changes at $\sim 0.8\%$ level for 195 kHz frequency separation at 60 MHz, which has a negligible contribution to the difference.} between two consecutive sub-bands 195 kHz apart, and any contribution due to the sky signal should largely drop out in the differential Stokes $I$ images. Similarly, differential Stokes $V$ images should contain only noise. However, effects which are non-smooth at the sub-band level are expected to leave their signature in the differential Stokes images.  

We use Stokes $I$ and $V$ residual images of the 3C220 field ($\nu_1 = 59.76$ MHz and $\nu_2 = 59.95$ MHz, located at the most sensitive part of the band) after DD-calibration to estimate the differential power spectra $P_{\Delta_{\nu} I}$ and $P_{\Delta_{\nu} V}$, and determine their ratio $P_{\Delta_{\nu} I}/P_{\Delta_{\nu} V}$. The top panel of figure \ref{fig:dPS_IV} shows $P_{\Delta_{\nu} I}$ (red solid curve) and $P_{\Delta_{\nu} V}$ (red dashed curve). We also show a slice of $P_{\Delta_{t} I}$ and $P_{\Delta_{t} V}$ at $59.95$ MHz in the same plot for comparison.  We observe that the power spectra are more or less flat on baselines $|\vect{u}| > 200\lambda$ and increase rapidly for $|\vect{u}| > 200\lambda$. This can be attributed to variations in the $uv$-coverage of LOFAR-LBA. The variations in these power spectra also correlate well with the $uv$-coverage profile shown in figure \ref{fig:uvcov}. 

The bottom panel of figure \ref{fig:dPS_IV} shows the ratio $P_{\Delta_{\nu} I}/P_{\Delta_{\nu} V}$ (red curve) and the ratio $P_{\Delta_{\nu} I}/P_{\Delta_{t} I}$ (blue curve). We also observe that the ratio $P_{\Delta_{\nu} I}/P_{\Delta_{\nu} V}$ is relatively flat and has value $\sim 2-3$ over the presented baseline range. This suggests that the rapid upturn in the power spectra shown in the top panel is due to $uv$-coverage variations and cancels out in the ratio. The excess variance in $P_{\Delta_{\nu} I}$ compared to $P_{\Delta_{\nu} V}$ is possibly due to random errors produced in calibration and/or erratic ionosphere. These random errors when applied to the data or the sky-model during subtraction, affect both Stokes $I$ and $V$. However, these errors lead to larger variance when applied to the emission in Stokes $I$ compared Stokes $V$ which lacks any emission (or below thermal noise, if any), resulting in excess noise at sub-band level. 

The ratio we observe here is considerably smaller than that we observed in G18 ($P_{\Delta_{\nu} I}/P_{\Delta_{\nu} V} \gtrsim 10$). This lower Stokes $I$ noise is in part achieved because \texttt{SAGEcal-CO} enforces frequency smoothness of the gain solutions in the calibration process, and also because the ionospheric activity is more benign compared to the observation in G18 where frequency smoothness was not enforced in the calibration and the ionosphere behaved erratically. To quantify the ionospheric activity, we use ionospheric Rotation Measure (RM) estimates from the GPS data. The ionospheric RM levels for the current observation are of order $\sim0.1-0.15\,\text{rad\,m}^{-2}$ during 90\% of the observation which is $\gtrsim 4$ times lower than the ionospheric RM levels (RM $> 0.4\,\text{rad\,m}^{-2}$) for the observation in G18, suggesting benign ionospheric activity.

Furthermore, from comparison of $P_{\Delta_{\nu} I}$ with $P_{\Delta_{t} I}$, we observe that there is a sudden jump in the ratio at $|\vect{u}|\sim 200\lambda$. The ratio is $\gtrsim 2$ for $|\vect{u}| < 200$ and it continues to increase as the baseline length decreases, compared to the values ($\sim 1-2$) for $|\vect{u}| > 200$. We attribute this effect to the $200\lambda$ baseline cut used in the DD-calibration. The sky-model incompleteness or ionospheric effects can introduce random errors during the calibration step. These random errors on gain solutions when applied to the baselines excluded during the calibration step, increase the variance in Stokes $I$ compared to Stokes $V$ on  excluded baselines \citep{patil2016,barry2016,ewall-wice2016,ewall-wice2017,gehlot2018,sardarabadi2018}. 

\section{Gaussian Process Regression}\label{sec:GPR}

After subtracting the calibration sky-model using DD-calibration, any remaining foreground emission within the primary beam consists of unresolved sources, sources below the confusion noise, sources not included in the model, and diffuse emission. These foregrounds should vary slowly with frequency, making them separable from the 21-cm signal which has rapid spectral fluctuations. We use a Gaussian Process Regression (GPR) technique (see \citealt{mertens2018} for more details) to remove this remaining foreground emission and other spectral structures imparted on the data due to instrumental mode-mixing, such as instrumental chromaticity and imperfect calibration residuals. In this section, we briefly describe GPR and its application to LOFAR-LBA data.

\begin{figure*}
\centering
\includegraphics[width=\textwidth]{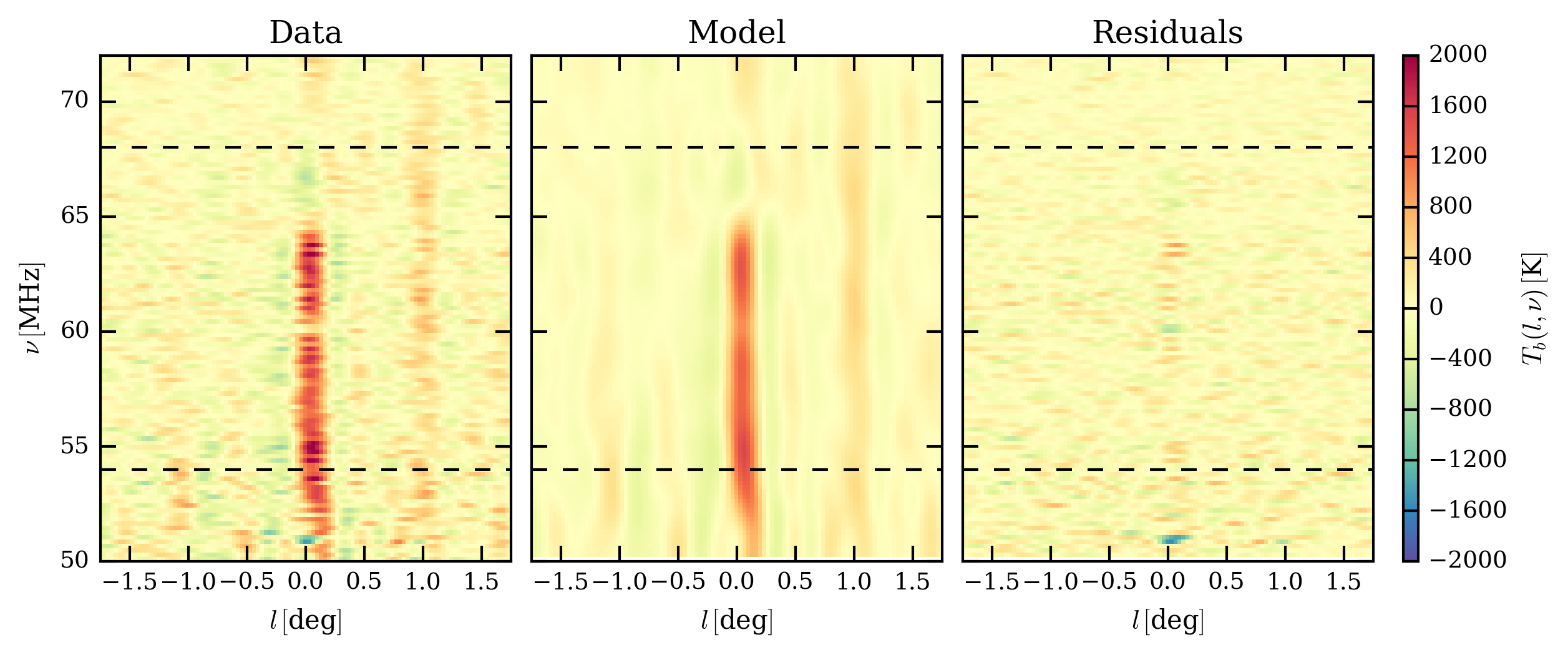}
    \caption{The 3C220 field Stokes $I$ image cube slices (in brightness temperature units) across the centre of a spatial axis after different processing steps. \textit{Left panel:} A slice of the image cube after DD-calibration (data). \textit{Middle panel:} The GPR model of the smooth foregrounds (intrinsic + mode-mixing) corresponding to the data. \textit{Right panel:} The residuals after subtracting the GPR model from the data. The dashed black lines represent the frequency range ($54-68$ MHz) used for power spectrum estimation. The residuals after GPR are featureless except for a few outliers.} 
\label{fig:3C220-foregrounds}
\end{figure*} 

\begin{figure*}
\centering
\includegraphics[width=\textwidth]{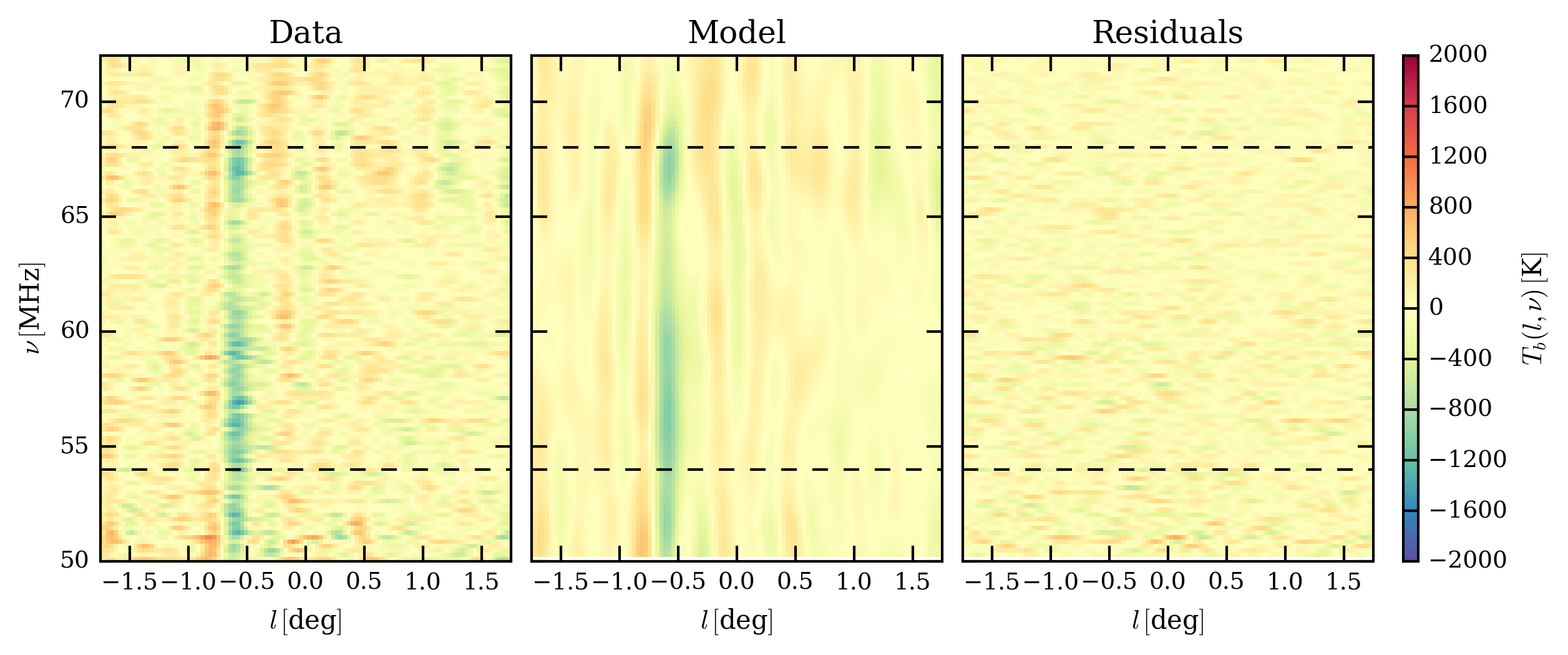}
    \caption{As figure \ref{fig:3C220-foregrounds} but for the NCP field. Similar to the 3C220 field, the residuals in the NCP field after GPR are featureless.} 
\label{fig:NCP-foregrounds}
\end{figure*} 

\subsection{Methodology}\label{subsec:GPR-method}

The visibilities observed by an interferometer ($\mathcal{V}_{\text{obs}}(\vect{u},\nu)$) can be written as a sum of different components viz. the signal of interest ($\mathcal{V}_{21}(\vect{u},\nu)$), the foreground contribution ($\mathcal{V}_{\text{sky}}(\vect{u},\nu)$), instrumental mode-mixing ($\mathcal{V}_{\text{mix}}(\vect{u},\nu)$) and noise ($\mathcal{V}_{\text{n}}(\vect{u},\nu)$), i.e.
\begin{equation}
\mathcal{V}_{\text{obs}}( \vect{u} ,\nu) = \mathcal{V}_{21}(\vect{u},\nu) + \mathcal{V}_{\text{sky}}(\vect{u},\nu) + \mathcal{V}_{\text{mix}}(\vect{u},\nu) + \mathcal{V}_{\text{n}}(\vect{u},\nu).
\end{equation}

Each of these components has a distinct spectral behaviour which is exploited by GPR to separate them from each other and eventually remove the foreground components from the observed visibilities, leaving residuals with the signal of interest buried below the noise \citep{mertens2018}. GPR models these different components with Gaussian Processes (GP). A GP ($f \sim \mathcal{GP} (m,\kappa)$) is the joint distribution of a collection of normally distributed random variables and is defined by its mean $m$ and covariance function $\kappa$. Values of $\kappa$ specify the covariance between pairs of points at different frequencies and determine the structure of the function (e.g. its smoothness in frequency) which can be modelled with a GP. GPs are often described by parameterized priors in GPR, and the GP prior is selected such that it maximises the Bayesian evidence, estimated by conditioning these priors to the observations (see \citealt{rasmussen2005} for an extensive review). The parameters of the covariance functions (also known as `hyper-parameters') can be estimated using standard optimisation or MCMC algorithms. The observed data ${\bf d}$, being a stacked set of gridded visibilities as a function of frequency, can be modelled as
\begin{equation}
{\bf d} = {\bf f}_{\text{fg}}(\nu) + {\bf f}_{21}(\nu) + {\bf n},
\end{equation}
where ${\bf f}_{\text{fg}}(\nu)$ corresponds to the foreground component, ${\bf f}_{21}(\nu)$ corresponds to the signal of interest and ${\bf n}$ is the noise. The 21-cm signal is expected to decorrelate over frequency scales > 1 MHz, whereas foregrounds are expected to be smooth on 1 MHz scales and decorrelate over a much larger frequency range. The covariance function $K\equiv \kappa$ for this model can be written as a sum of foreground covariance function $K_{\text{fg}}$ and a 21-cm signal covariance function $K_{21}$, i.e.
\begin{equation}
K = K_{\text{fg}} + K_{21}.
\end{equation}
$K_{\text{fg}}$ can further be decomposed into $K_{\text{int}}$, which corresponds to intrinsic foregrounds (large-scale correlation of $\sim 10 - 100$ MHz) and $K_{\text{mix}}$, which corresponds to instrumental mode mixing such as side-lobe noise (decorrelates within $\sim 1-5$ MHz). The joint probability distribution for the observed data ${\bf d}$ and function values ${\bf f}_{\text{fg}}$ of the foreground model at the same frequency $\nu$ is then given by
\begin{equation}
\begin{bmatrix} {\bf d} \\ {\bf f}_{\text{fg}} \end{bmatrix} \sim \mathcal{N} \left( \ \begin{bmatrix} 0 \\ 0 \end{bmatrix} , \begin{bmatrix} K_{\text{fg}} + K_{21} + \sigma_n^2 I & K_{\text{fg}} \\ K_{\text{fg}} & K_{\text{fg}} \end{bmatrix} \ \right),
\end{equation}
where $\sigma_n^2$ is the noise variance, $I$ is the identity matrix and $K\equiv K(\nu,\nu)$. After GPR, the foreground model can be retrieved as
\begin{gather}
E({\bf f}_{\text{fg}}) = K_{\text{fg}}\left[ K + \sigma_n^2 I \right]^{-1} {\bf d}, \\
\text{cov}({\bf f}_{\text{fg}}) = K_{\text{fg}} - K_{\text{fg}} \left[ K + \sigma_n^2 I \right]^{-1} K_{\text{fg}}, 
\end{gather}
where $E({\bf f}_{\text{fg}})$ and $\text{cov}({\bf f}_{\text{fg}})$ are the expectation values and covariance of the foregrounds respectively. The residuals ${\bf d}_{\text{res}}$ after foreground model subtraction are 
\begin{equation}
{\bf d}_{\text{res}} = {\bf d} - E({\bf f}_{\text{fg}}).
\end{equation}
Readers may refer to \cite{mertens2018} for a detailed description of the GPR technique and its application as a novel method for foreground removal and 21-cm signal estimation.

\subsection{Application of GPR to the LOFAR-LBA data}\label{subsec:GPR-application}

We use GPR to remove remaining foreground emission from the residual visibilities after DD calibration. We test various covariance functions as kernels to model different components of the residual visibilities in GPR. We use a Bayesian framework to compare different covariance functions and select those models that maximise the marginal likelihood (or Bayesian evidence). We tested several GPR covariance kernels e.g. Radial Basis Functions, Rational Quadratic functions and different classes of Matern covariance functions to model foreground components and finally selected the ones with the maximum Bayesian evidence.

To model the intrinsic foreground emission (unmodeled sources and diffuse emission) we choose a RBF(Radial Basis Function) covariance function as kernel. The RBF kernel is essentially a square exponential or Gaussian function defined as:
\begin{equation}
\kappa_{\text{RBF}} (\nu_{\text{p}},\nu_{\text{q}}) = \exp \left( \dfrac{-|\nu_{\text{q}} - \nu_{\text{p}}|^2}{2l^2} \right)
\end{equation}
where $l$ is the characteristic coherence scale in frequency. We use $5-100$ MHz as the prior for the range of coherence scales of the intrinsic foregrounds. To model the mode-mixing component of the foregrounds, we choose a Rational Quadratic (RQ) covariance function defined as:
\begin{equation}
\kappa_{\text{RQ}} (\nu_{\text{p}},\nu_{\text{q}}) = \left( 1 + \dfrac{|\nu_{\text{q}} - \nu_{\text{p}}|^2}{2\alpha l} \right)^{-\alpha} \ ,
\end{equation}
where $l$ is the coherence scale and $\alpha$ is the so-called power-parameter. RQ functions can be understood as infinite sums of Gaussian covariance functions with characteristic coherence scales \citep{rasmussen2005}. We use $1-8$ MHz as prior values for the coherence scales and $\alpha = 0.1$ for the mode-mixing component. To account for the 21-cm signal, we use an Exponential covariance function, which is a special case of a Matern class covariance function \citep{stein1999} defined as:
\begin{equation}
\kappa_{\text{matern}} (\nu_{\text{p}},\nu_{\text{q}}) = \dfrac{2^{1-n}}{\Gamma (n)} \left[ \dfrac{\sqrt{2n}|\nu|}{l} \right]^{n} K_{n} \left( \dfrac{\sqrt{2n}|\nu|}{l} \right) \ ,
\end{equation}
where $|\nu| = |\nu_{\text{q}} - \nu_{\text{p}}|$ and $K_{n}$ is the modified Bessel function of the second kind (not to be confused with the covariance functions defined in section \ref{subsec:GPR-method}). The `hyper-parameter' $l$ represents the characteristic coherence scale. Special classes of Matern covariance functions can be obtained by choosing various values for $n$, e.g. choosing $n = 1/2$ gives an exponential kernel. We use a frequency coherence scale of $0.01-1.5$ MHz for the 21-cm signal with an initial value of 0.5 MHz. Using \texttt{21cmFAST} simulations \citep{mesinger2007,mesinger2011}, \cite{mertens2018} have shown that these coherence scales covers a wide range of possible 21-cm signal models.

We use the residual image-cubes obtained after DD-calibration for foreground removal. We perform GPR foreground removal on the inner $3.5^{\circ}\times 3.5^{\circ}$ region of the image cubes (which is slightly smaller than the primary beam FWHM $\sim 4^{\circ}$) to limit sky curvature and primary beam effects. We selected the $50-72$ MHz frequency range for GPR foreground removal, which is 8 MHz wider than the power spectrum estimation window, for better foreground fitting and removal. Figure \ref{fig:3C220-foregrounds} shows slices through the Stokes $I$ image cubes for the 3C220 field across the center of one of the two spatial axes before GPR (left panel), the GPR foreground fit (middle panel) and the residuals after subtracting the foreground model obtained with GPR from the data (right panel). Similarly, figure \ref{fig:NCP-foregrounds} shows the slices of Stokes $I$ image cubes for the NCP field. We observe that the Stokes $I$ residuals after foreground removal with GPR for both the 3C220 and NCP fields are featureless (except for a few outliers) and do not appear to have spatial or spectral structure. In the following section, we use these residuals after GPR to create cylindrically and spherically averaged power spectra for the 3C220 and the NCP fields.

\begin{figure*}
\centering
\includegraphics[width=\textwidth]{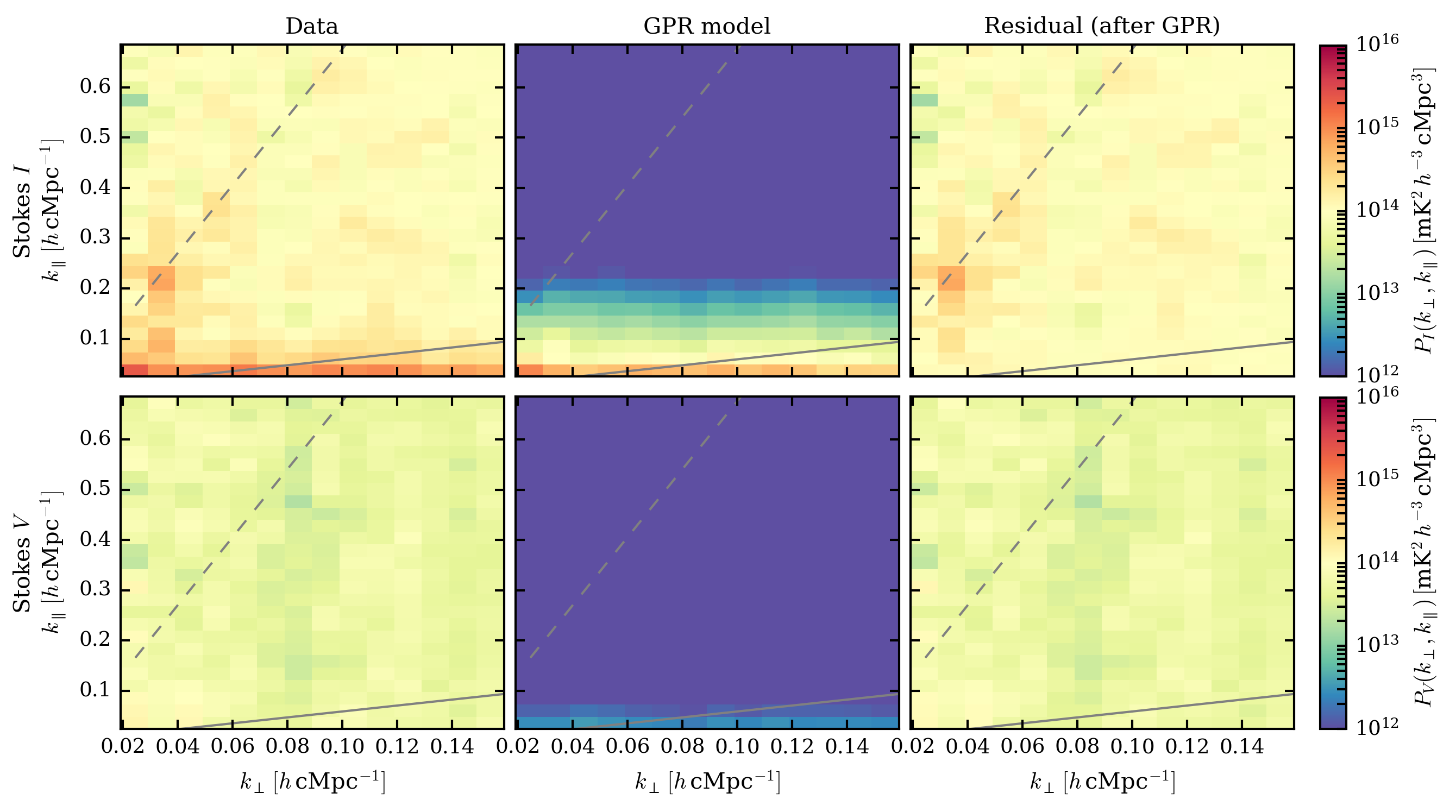}
    \caption{The cylindrically averaged Stokes $I$ and $V$ power spectra for the  3C220 field. \textit{Top row (left to right):} $P_{I}(k_{\perp},k_{\parallel})$ before foreground removal, GPR foreground model, and after foreground removal with. \textit{Bottom row (left to right):} Same as top row but for Stokes $V$. The solid grey lines correspond to a $5^{\circ}$ field of view which is slightly larger than the primary beam FWHM at 60 MHz. The dashed grey lines correspond to the instrumental horizon.} 
\label{fig:3C220-ps2d_IV}
\end{figure*} 

\begin{figure*}
\centering
\includegraphics[width=\textwidth]{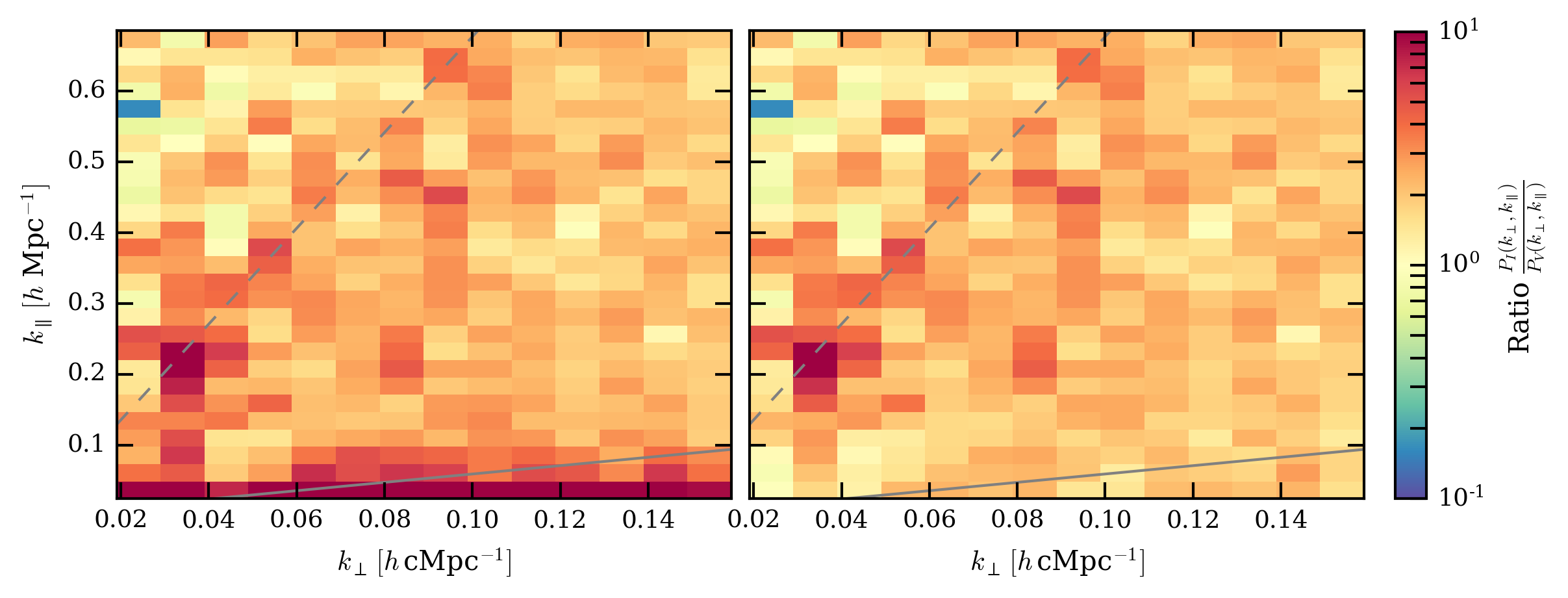}
    \caption{The ratio of the cylindrically averaged Stokes $I$ and $V$ power spectra for the 3C220 field. \textit{Left panel:} $P_{I}/P_{V}$ before foreground removal with GPR. \textit{Right panel:} $P_{I}/P_{V}$ after foreground removal with GPR.} 
\label{fig:3C220-ps2d_ratio}
\end{figure*} 

\section{Power Spectra Results}\label{sec:PSpec-results}

In this section we present the cylindrically and spherically averaged power spectra for the 3C220 and NCP fields. Cylindrically averaged power spectra (or 2D cosmological power spectra) are widely used in 21-cm experiments to assess various 21-cm signal contaminants such as foregrounds, side-lobe noise and systematic biases. Cylindrically averaged power spectra ($P(k_{\perp},k_{\parallel})$) are defined as \citep{parsons2012,thyagarajan2015a}:
\begin{equation}
P(k_{\perp},k_{\parallel}) = \dfrac{X^2 Y}{\Omega_A B} \langle | \tilde{\mathcal{V}}(\vect{u},\eta) |^2 \rangle ,
\end{equation}
where $\tilde{\mathcal{V}}(\vect{u},\eta)$ is the FT of the visibilities in the frequency direction, $\Omega_A$ is the integral of the square of the primary beam over solid angle across the sky, and $B$ is the bandwidth of the visibility cube. $X$ and $Y$ are the conversion factors from angle and frequency to transverse co-moving distance ($D(z)$) and the co-moving depth along the line of sight ($\Delta D$), respectively. The wave numbers $k_{\perp}$ and $k_{\parallel}$ are related to baseline vector ($\vect{u} = (u,v)$ in units of wavelength) and the Fourier dual to frequency ($\eta$) as:
\begin{equation}
k_{\perp} = \dfrac{2\upi \left( |\vect{u}| \right)}{D(z)} , \ \ k_{\parallel} = \dfrac{2\upi \nu_{21} H_{\text{0}} E(z)}{c(1+z)^2} \eta \ , 
\end{equation}
where $\nu_{21}$ is the rest frame frequency of the 21-cm spin flip transition of HI, $z$ is the redshift corresponding to the observation frequency, $c$ is the speed of light, $H_0$ is the Hubble constant and $E(z) \equiv \left[ \Omega_M(1+z)^3 + \Omega_k(1+z)^2 + \Omega_{\Lambda}\right]^{1/2} $ is a function of the standard cosmological parameters. The spherically averaged dimensionless power spectrum can be estimated from $P(k_{\perp},k_{\parallel})$ as:
\begin{equation}
\Delta^2(k) = \dfrac{k^3}{2\pi^2} P(k) ,
\end{equation}
where $k = \sqrt{k_{\perp}^2 + k_{\parallel}^2}$. We determine $P(k_{\perp},k_{\parallel})$ for both the 3C220 and NCP fields using the gridded visibility cubes of the $3.5^{\circ}\times 3.5^{\circ}$ region of the image cubes with 14 MHz bandwidth (54-68 MHz), corresponding to the redshift range $z=19.8 - 25.2$. We weigh the $uv$-cells using an empirical weighting scheme where $uv$-cells in Stokes $I$ and $V$ with higher noise are down-weighted. In this scheme, the weights are scaled by the inverse of the per-visibility variance $\sigma^2[\mathcal{V}(u,v)]$ in each $uv$-cell estimated by computing robust variance statistics of Stokes $V$ along the frequency direction as: 
\begin{equation}
\sigma^2[\mathcal{V}(u,v)] = \{ \hat{\sigma}_{\nu}[\sqrt{N_{\text{vis}}(u,v,\nu)} \times \mathcal{V}_{V}(u,v,\nu)]\} ^2    
\end{equation}
where $\mathcal{V}_{V}(u,v)$ are the gridded Stokes $V$ visibilities, $N_{\text{vis}}(u,v,\nu)$ is the number of visibilities in a $uv$-cell and $\hat{\sigma}_{\nu}$ is a robust standard deviation estimator. The weights ($W$) are then given by:
\begin{equation}
    W(u,v,\nu) = N_{\text{vis}}(u,v,\nu) \dfrac{\langle \sigma^2[\mathcal{V}(u,v)] \rangle}{\sigma^2[\mathcal{V}(u,v)]}.
\end{equation}
While the per-visibility variance should theoretically be identical in each $uv$-cell, it becomes different in the presence of systematics that can affect baselines differently. This empirical weighting scheme allows one to reduce the impact of those systematics. We emphasise that only Stokes $V$ is used in determining those weights. 

To Fourier transform the visibilities along the frequency direction, we use a Least Square Spectral Analysis (LSSA) method (full least squares FT-matrix inversion, see e.g. \citealt{barning1963,lomb1976,stoica2009,trott2016}). We apply a `Hann'\footnote{The `Hann' window is defined as $W(n) = 0.5 -0.5\cos \left[\dfrac{2\pi n}{(M-1)} \right]$, where $0\leq n\leq M-1$ (see e.g. \citealt{blackman1958,harris1978}).} window function to the data prior to the frequency transform to control the side-lobes along the $\eta$ axis, however, this window function somewhat increases the noise. Although using a `Hann' window introduces minor correlations between different $k_{\parallel}$ modes, the residual spectra are similar to the ones produced using a Top-hat window. Therefore, we currently ignore this effect in our analysis. The resulting $\tilde{\mathcal{V}}(\vect{u},\eta)$ cubes after frequency transform are scaled accordingly with the conversion factors $X$ and $Y$ calculated using $\Lambda$CDM cosmological parameters that are consistent with the Planck 2016 results \citep{planck2016}, and then cylindrically and spherically averaged to obtain $P(k_{\perp},k_{\parallel})$ and $\Delta^2(k)$, respectively. 

\subsection{The 3C220 field: cylindrical power spectra}\label{sec:PSpec_3C220}
In this section, we examine the cylindrical power spectra for the 3C220 field. The top row of figure \ref{fig:3C220-ps2d_IV} shows $P(k_{\perp},k_{\parallel})$ for Stokes $I$ before foreground removal, the GPR foreground model, and after GPR foreground removal. We observe that the lowest $k_{\parallel}$ bins in Stokes $I$ are dominated by smooth foreground emission (see top left panel) even after subtraction of the sky-model during DD-calibration. This foreground emission is modelled (shown in top middle panel) and effectively removed by the GPR foreground removal method (see top right panel). We also observe an excess power around the horizon line in Stokes $I$ prior to GPR, which is not removed during GPR, suggesting that this excess power has much lower spectral smoothness or decorrelates quickly over time between gridded visibilities and cannot be modelled with a GP with current settings. This structure is reminiscent of the `pitchfork' observed in G18 and is possibly caused by the residuals after Cas\,A and Cyg\,A subtraction. An inaccurate source model, ionospheric effects, the time variation of the primary beam, or a combination of these effects might explain these residuals. We expect the modelling errors to be negligible as their corresponding models are derived from high spatial resolution images and also because Cas\,A and Cyg\,A appear as compact sources on shorter baselines. Ionospheric effects, however, become stronger at lower elevations due to projection effects and subtraction of Cas\,A and Cyg\,A at 30 seconds and 61 kHz calibration resolution might not be sufficient to correct for ionospheric effects, especially on the shorter baselines. Also, the primary beam changes with time as the 3C220 field is tracked. Therefore, a combination of ionospheric effects, beam errors and time variation of the primary beam is likely capable of producing such an effect. 

The bottom row of figure \ref{fig:3C220-ps2d_IV} shows $P(k_{\perp},k_{\parallel})$ for Stokes $V$ before foreground removal, the GPR foreground model, and after GPR foreground removal. We observe that the Stokes V power spectrum is featureless before and after foreground removal, which suggests that any foreground emission/leakage in Stokes $V$ is lower than the excess variance in the current data (see bottom middle panel). We also do not observe any visible signature of Cas\,A and Cyg\,A residuals. The vertical bands in Stokes $I$ and $V$ near $k_{\perp} \approx 0.08$ and $0.14$ are due to varying $uv$-density and drop out in the ratio $\frac{P_{I}}{P_{V}}$. The ratio after foreground removal, as shown in the right panel of figure \ref{fig:3C220-ps2d_ratio}, is relatively flat compared to the one before foreground removal, except for the above-mentioned region near the horizon. The ratio has a median value of $2.07$ which is higher than the median value ($\sim 1.46$) observed in the ratio $P_{\Delta_t I} / P_{\Delta_t V}$ for the 3C220 field (see section \ref{subsec:ExcessNoise}). However, it is consistent with the excess at the sub-band level (see section \ref{subsec:excess_compare}) caused by the use of a baseline cut in the DD-calibration. 

\begin{figure*}
\centering
\includegraphics[width=1\textwidth]{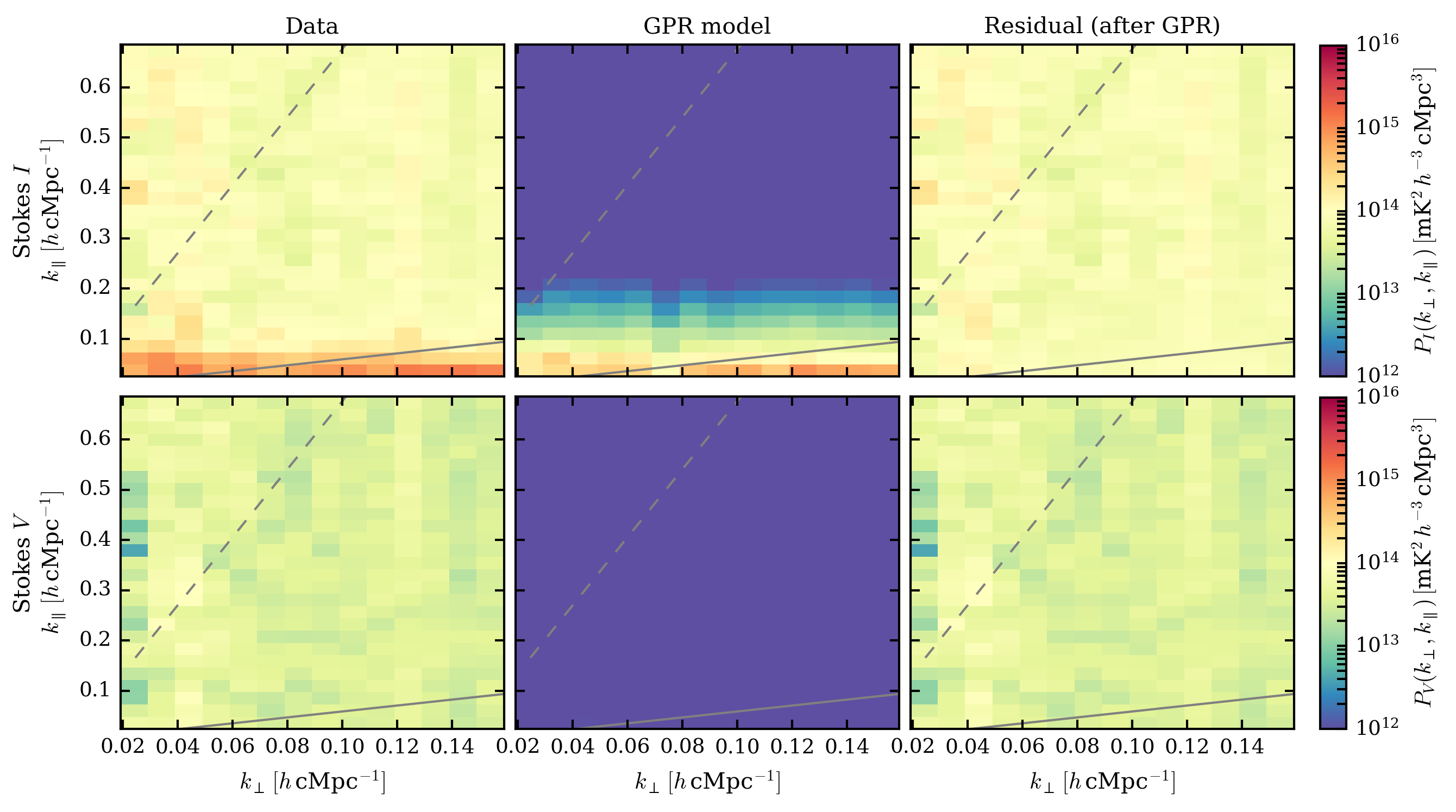}
    \caption{The cylindrically averaged Stokes $I$ and $V$ power spectra for the NCP field. \textit{Top row (left to right):} $P_{I}(k_{\perp},k_{\parallel})$ before foreground removal, GPR foreground model, and after foreground removal with. \textit{Bottom row (left to right):} Same as top row but for Stokes $V$.} 
\label{fig:NCP-ps2d_IV}
\end{figure*} 

\begin{figure*}
\centering
\includegraphics[width=\textwidth]{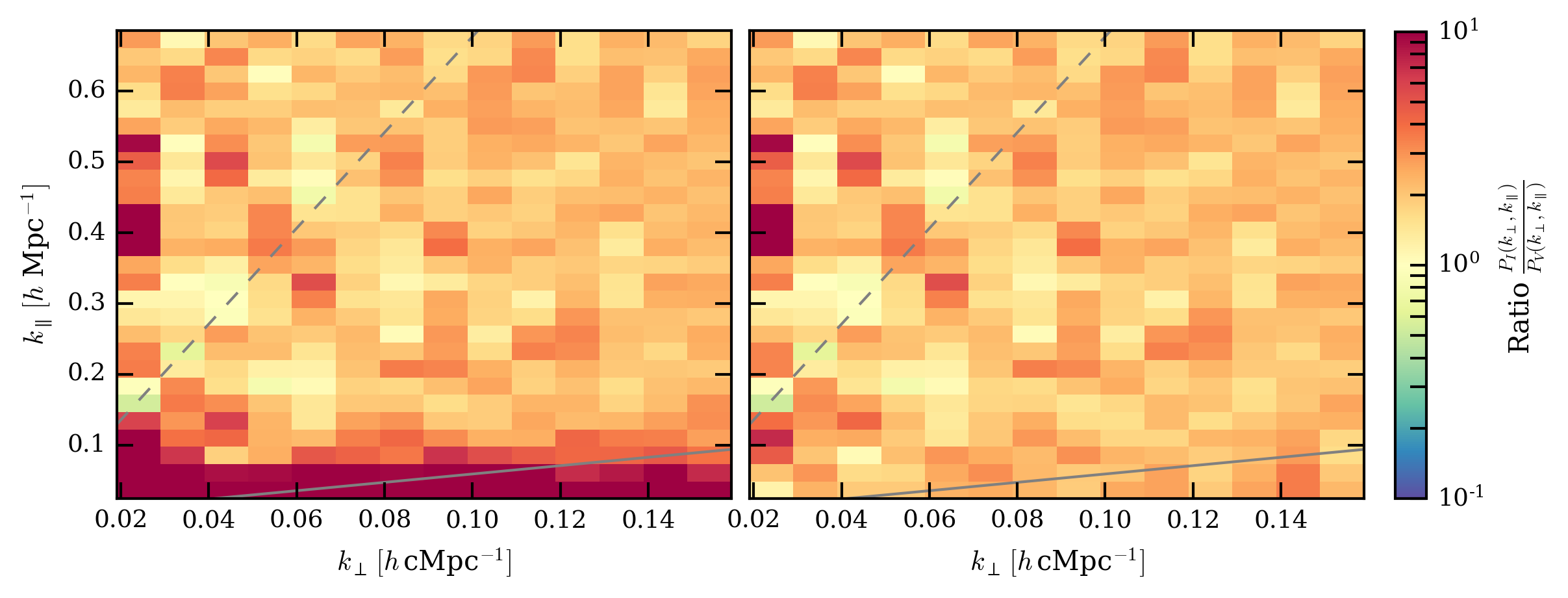}
    \caption{The ratio of the cylindrically averaged Stokes $I$ and $V$ power spectra for the NCP field. \textit{Left panel:} $P_{I}/P_{V}$ before foreground removal with GPR. \textit{Right panel:} $P_{I}/P_{V}$ after foreground removal with GPR.} 
\label{fig:NCP-ps2d_ratio}
\end{figure*} 

\subsection{The NCP field: cylindrical power spectra}\label{sec:PSpec_NCP}
In this section, we assess the cylindrical power spectra for the NCP field. Figure \ref{fig:NCP-ps2d_IV} shows $P(k_{\perp},k_{\parallel})$ for Stokes $I$ and $V$. The top left panel shows the power spectrum after DD-calibration, where low $k_{\parallel}$ modes are dominated by the power due to foreground emission. Similar to the 3C220 field, this power is effectively removed with GPR (see top right panel). We do not observe a `pitchfork' in Stokes $I$ or $V$ (presumably) due to Cas\,A and Cyg\,A residuals opposed to the 3C220 field. This might be primarily because the NCP is stationary on the sky and therefore the beam does not change (only rotates) as the observation progresses. It is also likely that the Cas\,A and Cyg\,A are closer to the null for the NCP, causing the power on/around the structure to be below the noise. Similar to the 3C220 field, Stokes $V$ power spectra for the NCP field appear featureless before and after foreground removal (see figure \ref{fig:NCP-ps2d_IV}). 

The behaviour of the ratio $\frac{P_{I}}{P_{V}}$ (see figure \ref{fig:NCP-ps2d_ratio}) is also similar to that of the 3C220 field. The ratio becomes relatively flat after foreground removal except for a few outliers at the small $k_{\perp}$ values. The ratio has a median value of $2.10$, which is almost equivalent to the median we observed for the 3C220 field, but higher than the median of the ratio $P_{\Delta_t I} / P_{\Delta_t V}$ for the NCP field (Section \ref{subsec:ExcessNoise}). This excess can also be attributed to the baseline cut in DD-calibration as discussed in Section \ref{subsec:excess_compare}, which we know causes excess power.

\begin{figure*}
\centering
\includegraphics[width=\textwidth]{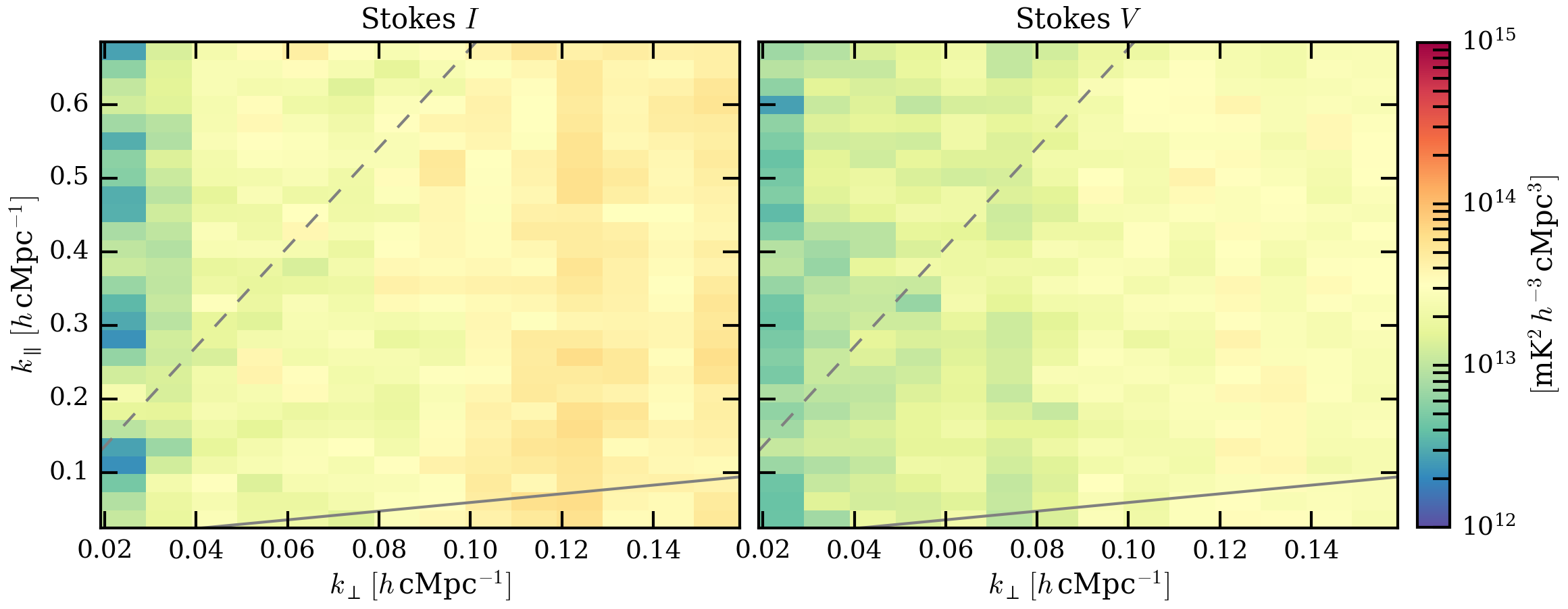}
    \caption{The cylindrically averaged Stokes $I$ and $V$ noise power spectra $P_{I}^N$ (left panel) and $P_{V}^N$ (right panel) for the 3C220 field determined from the difference cubes $\Delta_t\tilde{I}$ and $\Delta_t\tilde{V}$ respectively.} 
\label{fig:ps2d-noise_IV}
\end{figure*} 

\begin{figure}
\centering
\includegraphics[width=\columnwidth]{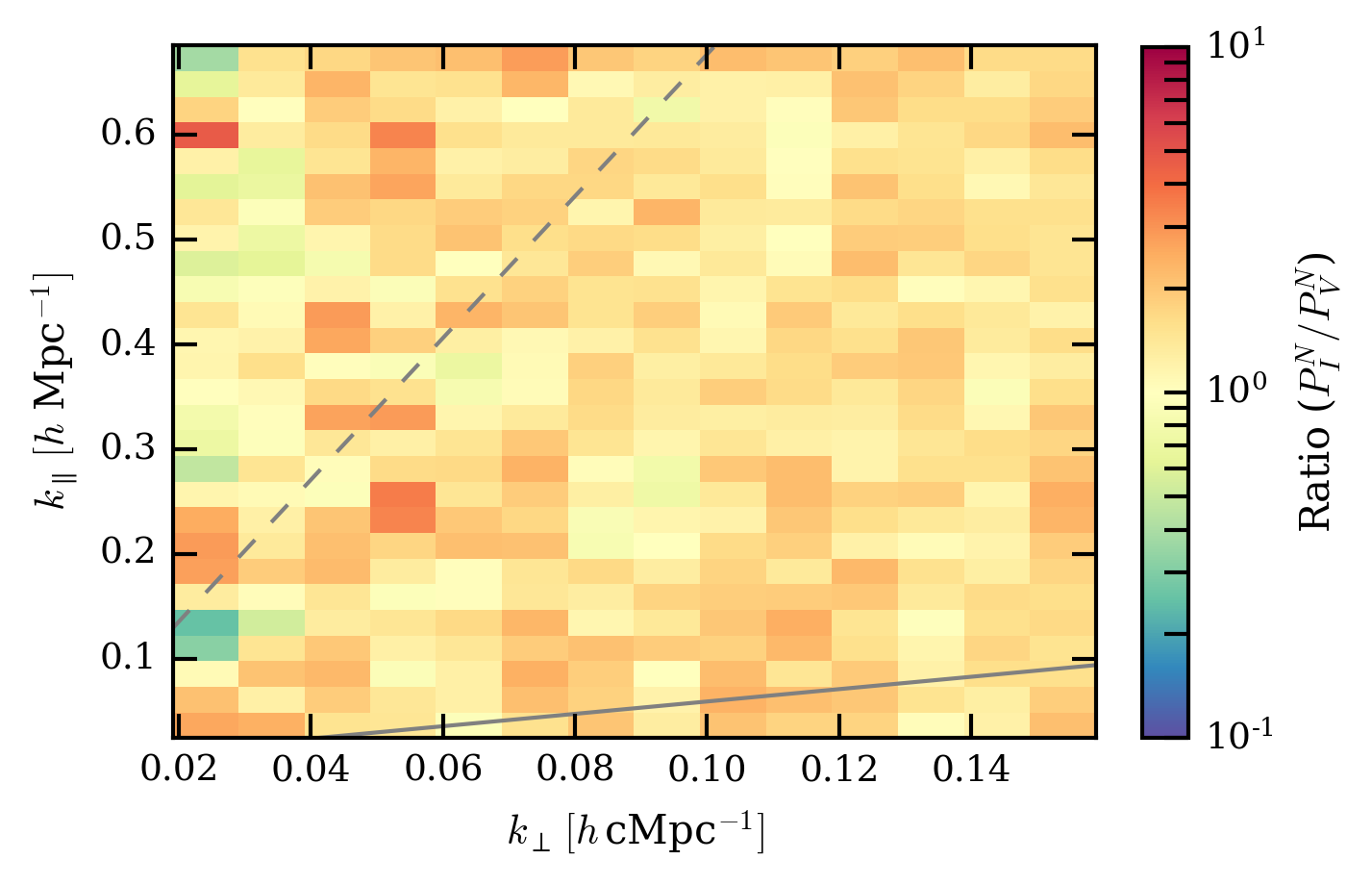}
    \caption{The ratio $\frac{P_{I}^N}{P_{V}^N}$ for the 3C220 field. We observe that the ratio has a median value of $1.51$.} 
\label{fig:ps2d-noise_ratio}
\end{figure} 

\subsection{Comparison with noise power spectra}\label{subsec:2Dcompare_with_noise}

We determine the cylindrically averaged noise power spectra $P_{I}^N$ and $P_{V}^N$ corresponding to the Stokes $I$ and $V$ difference cubes $\Delta_t \tilde{I}$ and $\Delta_t \tilde{V}$ respectively for the 3C220 field (e.g. see section \ref{sec:LBAnoise}), by passing these cubes through the power spectrum estimation pipeline. Note that we do not perform foreground removal on these data-cubes because we expect the sky component to drop out on time scales of 12 seconds. Figure \ref{fig:ps2d-noise_IV} shows $P_{I}^N$ (left panel) and $P_{V}^N$ (right panel). We observe that power in both Stokes $I$ and $V$ is lower for small $k_{\perp}$ values and higher for larger $k_{\perp}$ because the $uv$-density of LOFAR-LBA decreases with increasing baseline length and drops out in the ratio $\frac{P_{I}^N}{P_{V}^N}$ shown in figure \ref{fig:ps2d-noise_ratio}. From comparison of $P_V(k_{\perp},k_{\parallel})$ for the 3C220 and NCP fields with $P_{V}^N$, we notice that $P_V(k_{\perp},k_{\parallel})$ deviates from $P_{V}^N$ for lower $k_{\perp}$ ($<0.1$) values. This deviation in $P_V(k_{\perp},k_{\parallel})$ compared to $P_{V}^N$ can be attributed to the baseline cut used in the DD-calibration, which increases the noise on the baselines excluded from the calibration. 

Moreover, $\frac{P_{I}^N}{P_{V}^N}$ has a median value of $1.51$, which is consistent with the median value of $1.46$ for the ratio $P_{\Delta_t I} / P_{\Delta_t V}$. The NCP field, however, has a slightly lower median value of $1.3$. We observe that this excess power in Stokes $I$ for both the 3C220 and the NCP fields at 12 seconds level does not depend on the calibration, and is present at the same level throughout the analysis even after DD-calibration, foreground removal and also in the auto-correlations (results not shown here). This excess is different from the calibration cut induced noise and might have a physical origin. To account for this physical excess noise in the estimation of the spherically averaged power spectrum, we multiply the residual Stokes $V$ gridded visibilities after DD-calibration (since Stokes $V$ is an independent estimator of the thermal noise of the instrument) with the square-root of the median of the ratio $P_{\Delta_t I} / P_{\Delta_t V}$ (calculated in section \ref{subsec:ExcessNoise}) to obtain an estimate of the noise in Stokes $I$. We use the median instead of the mean because the median is a more robust representative of the central tendency of the distribution, whereas the mean is sensitive to outliers and becomes biased in the presence of strong outliers. This excess noise bias corrected Stokes $V$ is used as an estimator for the noise in the data in the foreground removal and spherically averaged power spectrum estimation steps. 

\begin{figure*}
\centering
\includegraphics[width=\textwidth]{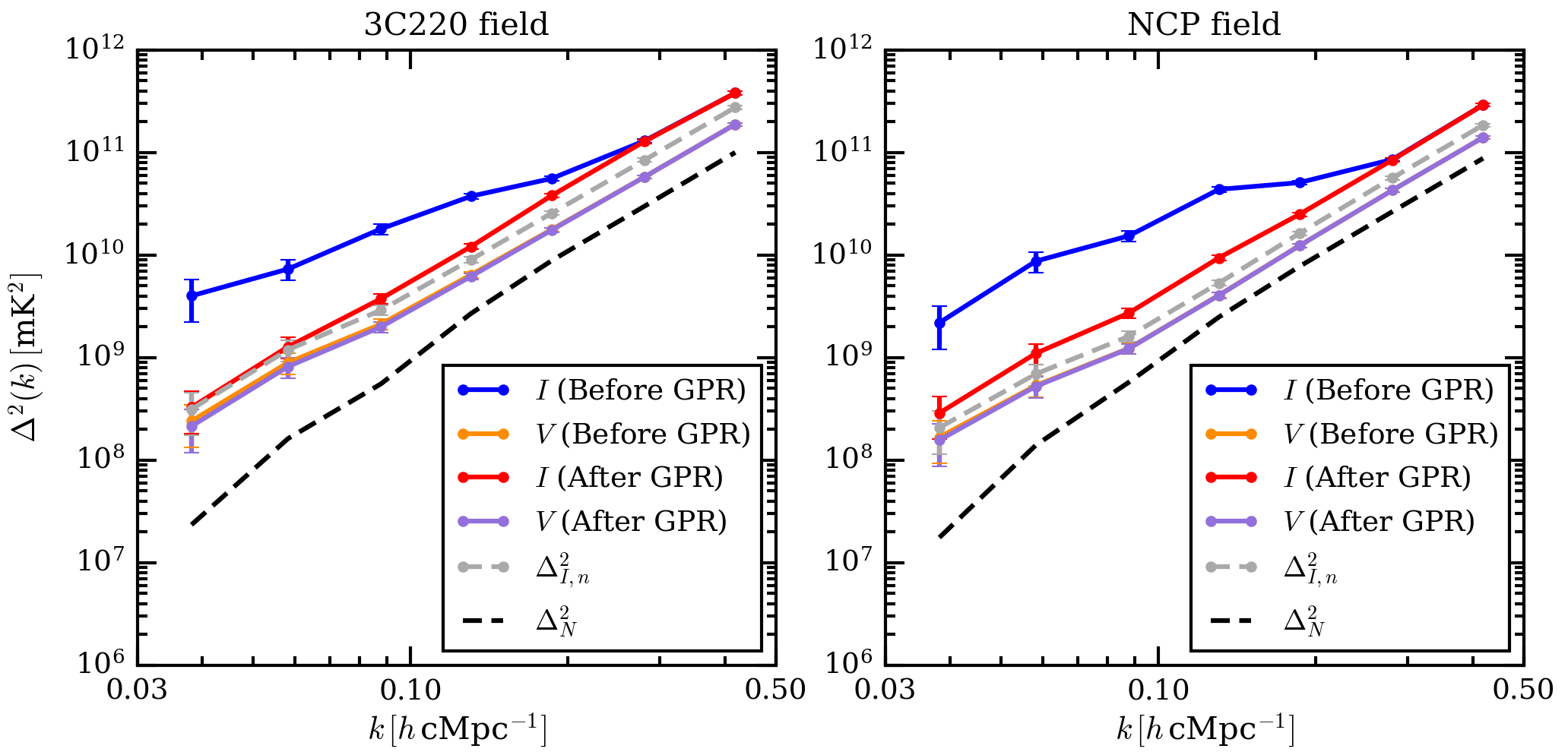}
    \caption{The spherically averaged Stokes $I$, $V$ and excess noise bias corrected Stokes $V$ power spectra. \textit{Left panel:} $\Delta_{I}^2$ and $\Delta_{V}^2$ for the 3C220 field before (blue and orange curves respectively) and after (red and purple curves respectively) foreground removal. \textit{Right panel}: $\Delta_{I}^2$ and $\Delta_{V}^2$ for the NCP field using the same colour scheme as in the left panel. The dashed grey and dashed black curves represent noise bias corrected Stokes $V$ power spectrum $\Delta_{I,n}^2$ and noise power spectrum estimate $\Delta_N^2$, respectively, for the corresponding fields. The errorbars represent the $2\sigma$ errors on the power spectra.} 
\label{fig:ps3d_IV}
\end{figure*} 
\begin{figure}
\centering
\includegraphics[width=\columnwidth]{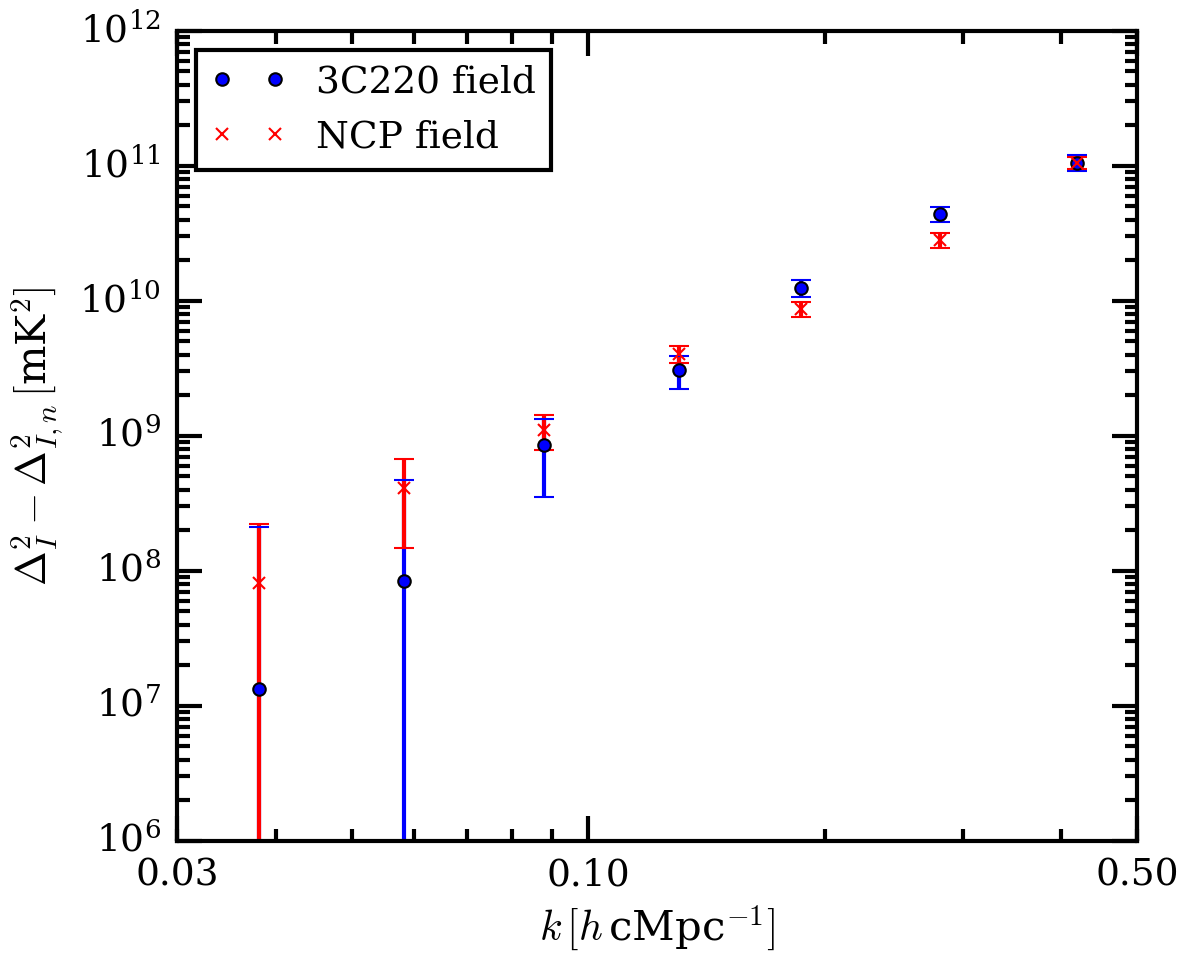}
    \caption{Noise bias corrected spherically averaged Stokes $I$ power spectra ($\Delta_I^2 - \Delta_{I,n}^2$) for the 3C220 and NCP fields. Blue circles represent the 3C220 field and red crosses represent the NCP field. The errorbars correspond to the $2\sigma$ errors on the power spectra.} 
\label{fig:ps3d_ulim}
\end{figure} 

\subsection{Spherically averaged power spectra}
We finally determine the Stokes $I$ and $V$ spherically averaged power spectra ($\Delta^2(k)$) for both the 3C220 and NCP fields before and after foreground removal for the redshift range $z = 19.8 - 25.2$. Figure \ref{fig:ps3d_IV} shows Stokes $I$ power spectra $\Delta_{I}^2$ and Stokes $V$ power spectra $\Delta_{V}^2$ for both the 3C220 (left panel) and NCP fields (right panel) before and after foreground removal. We use the physical excess noise bias corrected (using the median values from Section \ref{subsec:ExcessNoise}) Stokes $V$ visibilities as an estimator of the noise component in the Stokes $I$ power spectrum ($\Delta_{I,n}^2$), in order to account for the physical excess noise in Stokes $I$ compared to Stokes $V$. The dashed gray curves in figure \ref{fig:ps3d_IV} represent the excess bias corrected Stokes $V$ power spectra $\Delta_{I,n}^2$. For both fields, we observe that the power on smaller $k$ modes is dominated by large-scale foreground emission which comprises diffuse emission, unmodeled sources and sources below the confusion noise prior to foreground removal. A recent analysis of the wide-field AARTFAAC-12 HBA data (at 122 MHz) presented in \cite{gehlot2019} shows strong diffuse emission around the NCP on degree scales ($\vect{u} < 200$) which is well beyond the thermal noise. This emission becomes stronger at lower frequencies (50-70\,MHz) suggesting that the smallest $k$ modes are fully dominated by diffuse emission. Residual Stokes $I$ power on the smallest $k$ modes after foreground removal is an order of magnitude lower than the former. However, GPR does not remove any power from Stokes $V$, which means that any structure in Stokes $V$ is spatially and spectrally incoherent and behaves as uncorrelated noise. For the 3C220 field, Stokes $I$ residuals approach $\Delta_{I,n}^2$ at smaller $k$ modes, however these are still higher than $\Delta_{I,n}^2$ by $\sim 30\%$ on large $k$ modes. On the other hand, Stokes $I$ residuals for the NCP field are higher than $\Delta_{I,n}^2$ by $\sim50\%$ on most $k$ modes except for the lowest one. This remaining excess power, after correcting for the physical excess noise bias, is likely due to the baseline cut used during the DD-calibration.

Assuming that the physical noise properties of Stokes $I$ and $V$ are statistically identical, we use the post GPR excess noise bias corrected Stokes $V$ power spectrum ($\Delta_{I,n}^2$) to remove the noise component in the residual Stokes $I$ power spectrum. The noise bias corrected power spectrum $\Delta_I^2 - \Delta_{I,n}^2$ for the 3C220 (blue circles) and the NCP field (red crosses) are shown in figure \ref{fig:ps3d_ulim}. The dashed curves show thermal noise power spectrum estimate $\Delta^2_N$ estimated from $\Delta_t\tilde{V}$ for the 3C220 (`skyblue' coloured) and the NCP field (`coral' coloured). We observe that $\Delta_I^2 - \Delta_{I,n}^2$ for both fields are consistent with each other within the $2\sigma$ uncertainty for modes $k\lesssim 0.2\,h\,\text{cMpc}^{-1}$ and deviate slightly on $k\gtrsim 0.2\,h\,\text{cMpc}^{-1}$.  This is possibly due to different morphologies of the two fields on small angular scales. The $\Delta_I^2 - \Delta_{I,n}^2$ for both fields, within $2\sigma$ uncertainty, agree with their respective noise estimate $\Delta_N^2$ (determined from $\Delta_t\tilde{V}$) which is a more accurate estimator of the thermal noise of the system. The $\Delta_N^2$ for both fields show power-law like behaviour and agree with each other on all $k$ modes. We find a $2\sigma$ upper limit of $\Delta_{21}^2 < (14561\,\text{mK})^2$ at $k\sim 0.038\,h\,\text{cMpc}^{-1}$ for the 3C220 field and $\Delta_{21}^2 < (14886\,\text{mK})^2$ at $k\sim 0.038\,h\,\text{cMpc}^{-1}$ for the NCP field in the redshift range $z = 19.8-25.2$. Both upper limits are consistent with each other within $2\%$. The upper limits $\Delta_I^2 - \Delta_{I,n}^2$ for the two fields are still dominated by systematics on most $k$ modes. A deeper understanding of systematics (and their mitigation) and a more accurate estimate of the noise bias is required to remove this additional bias. From our current analysis, we observed that it is harder to model the exact noise bias, which is crucial to obtain more robust upper limit. We are currently developing improved estimators of the incoherent noise power spectrum which might be thermal noise and also include incoherent random errors e.g. due to the ionosphere, calibration etc for noise bias subtraction. We are also exploring other methods (e.g. cross-correlating independent datasets) to estimate 21-cm power spectrum which circumvents several issues with noise bias subtraction and plan to incorporate them in future analyses.

\section{Summary and Outlook}\label{sec:conclusions}

In this work, we have explored the possibility of statistical measurement of the redshifted 21-cm signal of neutral hydrogen from the Cosmic Dawn using the LOFAR-Low Band Antenna system. We have presented the first upper limits on the power spectrum of the 21-cm signal in the high redshift range of $z = 19.8 - 25.2$ using LOFAR-LBA data with dual-pointing setup pointed at the NCP and the  radio galaxy 3C220.3 simultaneously. Our main conclusions are:
\begin{enumerate}

\item For the 3C220 field, after 14 hours of integration, a $2\sigma$ upper limit of $\Delta_{21}^2 < (14561 \ \text{mK)}^2$ at $k = 0.038\,h\,\text{cMpc}^{-1}$ is reached on the power spectrum of 21-cm brightness temperature fluctuations. Similarly, for the NCP field, we reach a $2\sigma$ upper limit of $\Delta_{21}^2 < (14886 \ \text{mK)}^2$ at $k = 0.038 \,h\,\text{cMpc}^{-1}$ in the redshift range $z = 19.8 - 25.2$. Both upper limits are consistent with each other within $2\%$ level. Upper limits for both the 3C220 and the NCP fields are still dominated by the systematics.

\item We demonstrate the application of a multiple pointing method to calibrate LOFAR-LBA dual pointing observations. 

\item We observe an excess of noise in the ratio of Stokes $I$ and $V$ noise spectra over short time-scales (12 seconds) in baseline-frequency space, derived from the Stokes $I$ and $V$ difference image-cubes created from even and odd visibility samplings at 12-second level. This excess is independent of frequency and baseline length and is also not affected by calibration. This excess noise is different from that introduced during calibration and already exists before DI and after DD calibration and does not change during those steps. The excess is different for the two fields and seems to have no clear origin. We suspect it to be caused by (diffractive) ionospheric scintillation noise, but we leave this analysis for the future.   

\item We show that introducing frequency smoothness of instrumental gains as a constraint in both Direction Independent and Direction Dependent calibration of LOFAR-LBA data greatly reduces the calibration induced excess variance on the sub-band level in Stokes $I$ compared to Stokes $V$ in contrast to the un-constrained case presented in G18, where we found an excess by a factor $\sim 10$. However, an excess of $\sim 2 - 3$ still remains, which can be explained by the exclusion of short baselines during DD-calibration as shown for LOFAR-HBA data calibration as well in \cite{patil2016} and \cite{sardarabadi2018}. 

\item After foreground removal using Gaussian Process Regression, the Stokes $I$ power spectrum is $\sim 2$ times that of Stokes $V$ for both fields and is featureless on most scales. However, we observe a `pitchfork' like structure in the 3C220 field at low $k_{\perp}$ near the horizon line. We expect this structure to be caused by Cas\,A and Cyg\,A residuals as seen by G18.

\end{enumerate}

\subsection{Outlook}

Detection of the redshifted 21-cm HI signal from Cosmic Dawn and the Epoch of Reionization promises to be an excellent probe to study the early phases of the evolution of the universe and has the potential to unveil exotic astrophysical phenomena. With the analysis shown in this work, a CD experiment with LOFAR-LBA will require $>10^4$ hours of integration (power spectrum sensitivity of $\sim(100\,\text{mK})^2$ in CD redshift range) in order to constrain the optimistic CD X-ray heating and baryon-Dark Matter scattering models (see e.g \citealt{fialkov2018,cohen2018}). We plan to improve the analysis in the future by improving the enforcement of spectral smoothness in calibration, better modelling of the instrument (improving beam models) and by using the new Image Domain Gridder (IDG) combined with \texttt{WSClean} (see e.g. \citealt{veenboer2017,vandertol2018}) to subtract off-axis sources. The upcoming LOFAR 2.0 upgrade will also increase the sensitivity of the LOFAR-LBA system. The combination of all these improvements will allow us to improve the CD 21-cm power spectrum sensitivity significantly.

Moreover, recently a deep absorption feature ($-0.5$ K deep) centred at $\sim78$ MHz ($z \sim 17 $) in the averaged sky spectrum was presented by \cite{bowman2018}, possibly being the sought-after 21-cm signal absorption feature seen against the Cosmic Microwave Background during the CD era. The suggested absorption feature is considerably ($\sim 2.5$ times) stronger and wider than predicted by standard astrophysical models \citep{barkana2018}. If confirmed, such a strong signal will lead to a large increase in the 21-cm brightness temperature fluctuations in the redshift range $z = 17-19$ corresponding to the deepest part of the absorption profile \citep{barkana2018,fialkov2018}, making it possible to detect its signal in a much shorter integration time compared to what was previously expected. Motivated by this, we have commenced a large scale program called the AARTFAAC Cosmic Explorer (ACE) which uses the Amsterdam-ASTRON Radio Transients Facility And Analysis Center (AARTFAAC) correlator based on LOFAR, to obtain wide-field data for statistical detection of the 21-cm brightness temperature fluctuations within the redshift range of the absorption feature. The techniques discussed in this paper and lessons learned here will be useful in understanding and mitigating the challenges in AARTFAAC data processing and analysis, as well as in the NenuFAR, the HERA and the upcoming SKA-low, which can also observe the similar redshift range. The SKA-low will also support multi-beam observations, and thus also can take advantage of the dual-beam calibration strategy we have demonstrated.

\section*{Acknowledgements}

BKG and LVEK acknowledge the financial support from a NOVA cross-network grant. FGM acknowledges support from a SKA-NL roadmap grant from the Dutch Ministry of OCW. LOFAR, the Low Frequency Array designed and constructed by ASTRON, has facilities in several countries, that are owned by various parties (each with their own funding sources), and that are collectively operated by the International LOFAR Telescope (ILT) foundation under a joint scientific policy.



\bibliographystyle{mnras}
\bibliography{bibentries.bib} % if your bibtex file is called example.bib

% Don't change these lines
\bsp	% typesetting comment
\label{lastpage}
\end{document}